\documentclass[zpreprint,zbstpl]{macros/zeus_paper}

\usepackage[english]{babel}

\chardef\usc=95
\chardef\til=126
\catcode`\@=11 
\DeclareRobustCommand\xdotspace{\futurelet\@let@token\@xdotspace}
\def\@xdotspace{%
  \ifx\@let@token.\else
  \ifx\@let@token\bgroup.\else
  \ifx\@let@token\egroup.\else
  \ifx\@let@token\/.\else
  \ifx\@let@token\ .\else
  \ifx\@let@token~.\else
  \ifx\@let@token!.\else
  \ifx\@let@token,.\else
  \ifx\@let@token:.\else
  \ifx\@let@token;.\else
  \ifx\@let@token?.\else
  \ifx\@let@token/.\else
  \ifx\@let@token'.\else
  \ifx\@let@token).\else
  \ifx\@let@token-.\else
  \ifx\@let@token\@xobeysp.\else
  \ifx\@let@token\space.\else
  \ifx\@let@token\@sptoken.\else
   .\space
   \fi\fi\fi\fi\fi\fi\fi\fi\fi\fi\fi\fi\fi\fi\fi\fi\fi\fi}
\catcode`\@=12 

\newcommand{\stru}[2]{%
   \relax\ifmmode\hbox{\vrule height#1 depth#2 width0pt}%
   \else\vrule height#1 depth#2 width0pt\fi}

\newcommand{\Ronum}[1]{\uppercase\expandafter{\romannumeral#1}}
\newcommand{\ronum}[1]{\expandafter{\romannumeral#1}}
\DeclareRobustCommand{\LaTeXZ}{%
  \LaTeX\kern-.05em4\kern-.1em
  {\raisebox{-0.2ex}{$\scriptstyle\text{ZEUS}$}}\xspace}

\DeclareMathAlphabet{\mathbf}{OT1}{cmr}{bx}{sl}
\newcommand{\eVdist}{\kern-0.06667em}

\newcommand{\pb}{\,\text{pb}}

\newcommand{\kg}{\,\text{kg}}

\newcommand{\slashfrac}[2]{%
  \raisebox{0.5ex}{\ensuremath #1}\kern-0.12em/\kern-0.08em
  \raisebox{-.8ex}{\ensuremath #2}}

\newcommand{\sqr}[3]{%
    {\vcenter{\hrule height.#3ex\hbox{\vrule width.#2ex height#1ex
     \kern#1ex\vrule width.#3ex}\hrule height.#2ex}}}

\catcode`\@=11 
\newcommand{\parenbar}{\mathpalette\p@renb@r}
\def\p@renb@r#1#2{\vbox{%
  \ifx#1\scriptscriptstyle \dimen@.7em\dimen@ii.2em\else
  \ifx#1\scriptstyle \dimen@.8em\dimen@ii.25em\else
  \dimen@1em\dimen@ii.4em\fi\fi \offinterlineskip
  \ialign{\hfill##\hfill\cr
    \vbox{\hrule width\dimen@ii}\cr
    \noalign{\vskip-.3ex}%
    \hbox to\dimen@{$\mathchar300\hfil\mathchar301$}\cr
    \noalign{\vskip-.3ex}%
    $#1#2$\cr}}}
\catcode`\@=12 

\newcommand{\IP}{{\rm I$\kern-0.01667em$P}\xspace}

\mathchardef\qsm=63
\mathchardef\pls=43
\mathchardef\mns=512
\mathchardef\plm=518
\mathchardef\eql=61
\mathchardef\smallleft=300
\mathchardef\smallright=301
\mathchardef\les=316
\mathchardef\gre=318
\mathchardef\leq=532
\mathchardef\grq=533

\catcode`\@=11 
\newcounter{pict@width}
\newcounter{pict@height}
\newlength{\pict@scale}
\newcommand{\psfigadd}[4]{%
\setcounter{pict@width}{1*\ratio{#2+\pict@scale/2}{\pict@scale}}
\setcounter{pict@height}{1*\ratio{#3+\pict@scale/2}{\pict@scale}}
\setlength{\unitlength}{\pict@scale}
\hbox to #2{\hspace{-\fill}\begin{picture}(\thepict@width,\thepict@height)
\put(0,0){\psfig{figure=#1,width=#2,height=#3,clip=}}
\SetScale{0.283466457}
\SetWidth{1.763889}
{#4}
\end{picture}}
}
\newcounter{pict@widthfst}
\newcounter{pict@widthscd}
\newcounter{pict@widthtot}
\newcommand{\psfigaddtwo}[7]{%
\setcounter{pict@widthfst}{1*\ratio{#2+\pict@scale/2}{\pict@scale}}
\setcounter{pict@widthscd}{1*\ratio{#2+#4+\pict@scale/2}{\pict@scale}}
\setcounter{pict@widthtot}{1*\ratio{#2+#4+#6+\pict@scale/2}{\pict@scale}}
\setcounter{pict@height}{1*\ratio{#3+\pict@scale/2}{\pict@scale}}
\setlength{\unitlength}{\pict@scale}
\hbox{\hspace{-\fill}\begin{picture}(\thepict@widthtot,\thepict@height)
\put(0,0){\psfig{figure=#1,width=#2,height=#3,clip=}}
\put(\thepict@widthscd,0){\psfig{figure=#5,width=#6,height=#3,clip=}}
\SetScale{0.283466457}
\SetWidth{1.763889}
{#7}
\end{picture}}
}
\newcommand{\psfigror}[4]{%
\setcounter{pict@width}{1*\ratio{#2+\pict@scale/2}{\pict@scale}}
\setcounter{pict@height}{1*\ratio{#3+\pict@scale/2}{\pict@scale}}
\setlength{\unitlength}{\pict@scale}
\hbox{\begin{picture}(\thepict@width,\thepict@height)
\put(0,\thepict@height){\psfig{figure=#1,width=#3,height=#2,clip=,angle=270}}
\SetScale{0.283466457}
\SetWidth{1.763889}
{#4}
\end{picture}}
}
\newcommand{\psfigrol}[4]{%
\setcounter{pict@width}{1*\ratio{#2+\pict@scale/2}{\pict@scale}}
\setcounter{pict@height}{1*\ratio{#3+\pict@scale/2}{\pict@scale}}
\setlength{\unitlength}{\pict@scale}
\hbox{\begin{picture}(\thepict@width,\thepict@height)
\put(0,0){\psfig{figure=#1,width=#3,height=#2,clip=,angle=90}}
\SetScale{0.283466457}
\SetWidth{1.763889}
{#4}
\end{picture}}
}
\catcode`\@=12 
\newlength\listtextwidth

\catcode`\@=11 
\newlength{\@tabfninsert}
\newlength{\@tabfnwidth}
\newcommand{\tabfootnote}[2]{%
  \setlength{\@tabfninsert}{0.8em}
  \setlength{\@tabfnwidth}{\textwidth}
  \addtolength{\@tabfnwidth}{-\@tabfninsert}
  \addtolength{\@tabfnwidth}{-0.4em}
  \noindent\makebox[\@tabfninsert][r]{\footnotesize$^{#1}$\hfil}\hfill%
  \parbox[t]{\@tabfnwidth}{\footnotesize #2\hfill}}
\catcode`\@=12 

\def\etjet{E_T^{\rm jet}}
\def\etajet{\eta^{\rm jet}}
\def\phijet{\varphi^{\rm jet}}
\def\etaphi{\eta-\varphi}
\def\etar{-1<\etajet<2.5}
\def\q2{Q^2}
\def\pb1{pb$^{-1}$}
\def\g2{GeV$^2$}
\def\mj{M^{\rm jj}}
\def\m3j{M^{\rm 3j}}
\def\kt{k_T}
\def\ele{e^+e^-}
\def\pp{p\bar p}
\def\colab#1{#1 Collaboration}
\def\z0{Z^0}
\def\mz{M_Z}
\def\mw{M_W}
\def\as{\alpha_s}
\def\asz{\as(\mz)}
\def\etal{et al.}
\def\ept{ep\rightarrow e t X}

\def\kg{\kappa_{tu\gamma}}

\def\vz{v_{tuZ}}
\def\mt{M_{\rm top}}

\begin{document}

\prepnum{{DESY--03--012}}

\title{
Search for single-top production in {\boldmath $ep$} collisions at HERA}
                    
\author{ZEUS Collaboration}
\date{January 2003}

\abstract{
A search for single-top production, $\ept$, has been made with the
ZEUS detector at HERA using an integrated luminosity of $130.1$~\pb1.
Events from both the leptonic and hadronic decay channels of the $W$ boson
resulting from the decay of the top quark were sought.
For the leptonic mode, the search was made for events with
isolated high-energy leptons and significant missing transverse
momentum. For the hadronic decay mode, three-jet events
in which two of the jets had an invariant mass consistent with that of the
$W$ were selected. No evidence for top production was found. 
The results are used to constrain single-top
production via flavour-changing neutral current (FCNC) transitions.
The ZEUS limit excludes a substantial region in the FCNC $tu\gamma$
coupling not ruled out by other experiments.
}

\makezeustitle

\def\3{\ss}                                                                                        

\pagenumbering{Roman}                                                                              

\begin{center}                                                                                     
{                      \Large  The ZEUS Collaboration              }                               
\end{center}                                                                                       
  S.~Chekanov,                                                                                     
  M.~Derrick,                                                                                      
  D.~Krakauer,                                                                                     
  J.H.~Loizides$^{   1}$,                                                                          
  S.~Magill,                                                                                       
  B.~Musgrave,                                                                                     
  J.~Repond,                                                                                       
  R.~Yoshida\\                                                                                     
 {\it Argonne National Laboratory, Argonne, Illinois 60439-4815}~$^{n}$                            
\par \filbreak                                                                                     
  M.C.K.~Mattingly \\                                                                              
 {\it Andrews University, Berrien Springs, Michigan 49104-0380}                                    
\par \filbreak                                                                                     
  P.~Antonioli,                                                                                    
  G.~Bari,                                                                                         
  M.~Basile,                                                                                       
  L.~Bellagamba,                                                                                   
  D.~Boscherini,                                                                                   
  A.~Bruni,                                                                                        
  G.~Bruni,                                                                                        
  G.~Cara~Romeo,                                                                                   
  L.~Cifarelli,                                                                                    
  F.~Cindolo,                                                                                      
  A.~Contin,                                                                                       
  M.~Corradi,                                                                                      
  S.~De~Pasquale,                                                                                  
  P.~Giusti,                                                                                       
  G.~Iacobucci,                                                                                    
  A.~Margotti,                                                                                     
  R.~Nania,                                                                                        
  F.~Palmonari,                                                                                    
  A.~Pesci,                                                                                        
  G.~Sartorelli,                                                                                   
  A.~Zichichi  \\                                                                                  
  {\it University and INFN Bologna, Bologna, Italy}~$^{e}$
% ...
\newpage                                                                                           
$^{\    1}$ also affiliated with University College London \\                                      
% ...
                                                           %
                                                           %
\newpage   
                                                           %
                                                           %
\begin{tabular}[h]{rp{14cm}}                                                                       
$^{a}$ &  supported by the Natural Sciences and Engineering Research                               
          Council of Canada (NSERC) \\                                                             
$^{b}$ &  supported by the German Federal Ministry for Education and                               
          Research (BMBF), under contract numbers HZ1GUA 2, HZ1GUB 0, HZ1PDA 5, HZ1VFA 5\\         
$^{c}$ &  supported by the MINERVA Gesellschaft f\"ur Forschung GmbH, the                          
          Israel Science Foundation, the U.S.-Israel Binational Science                            
          Foundation and the Benozyio Center                                                       
          for High Energy Physics\\                                                                
$^{d}$ &  supported by the German-Israeli Foundation and the Israel Science                        
          Foundation\\                                                                             
$^{e}$ &  supported by the Italian National Institute for Nuclear Physics (INFN) \\                
$^{f}$ &  supported by the Japanese Ministry of Education, Science and                             
          Culture (the Monbusho) and its grants for Scientific Research\\                          
$^{g}$ &  supported by the Korean Ministry of Education and Korea Science                          
          and Engineering Foundation\\                                                             
$^{h}$ &  supported by the Netherlands Foundation for Research on Matter (FOM)\\                   
$^{i}$ &  supported by the Polish State Committee for Scientific Research,                         
          grant no. 620/E-77/SPUB-M/DESY/P-03/DZ 247/2000-2002\\                                   
$^{j}$ &  partially supported by the German Federal Ministry for Education                         
          and Research (BMBF)\\                                                                    
$^{k}$ &  supported by the Fund for Fundamental Research of Russian Ministry                       
          for Science and Edu\-cation and by the German Federal Ministry for                       
          Education and Research (BMBF)\\                                                          
$^{l}$ &  supported by the Spanish Ministry of Education and Science                               
          through funds provided by CICYT\\                                                        
$^{m}$ &  supported by the Particle Physics and Astronomy Research Council, UK\\                   
$^{n}$ &  supported by the US Department of Energy\\                                               
$^{o}$ &  supported by the US National Science Foundation\\                                        
$^{p}$ &  supported by the Polish State Committee for Scientific Research,                         
          grant no. 112/E-356/SPUB-M/DESY/P-03/DZ 301/2000-2002, 2 P03B 13922\\                    
$^{q}$ &  supported by the Polish State Committee for Scientific Research,                         
          grant no. 115/E-343/SPUB-M/DESY/P-03/DZ 121/2001-2002, 2 P03B 07022\\                    
\end{tabular}                                                                                      

\newpage
\pagenumbering{arabic} 
\pagestyle{plain}

\section{Introduction}
\label{secintro}
The Standard Model (SM) of the fundamental interactions presently
provides an accurate description of the phenomena observed in
both low- and high-energy reactions of elementary particles. As probes
in search for physics beyond the SM, observables sensitive to
flavour-changing neutral current (FCNC) interactions are particularly
useful, since the SM rates are very small due to the GIM
mechanism~\cite{gim}. The FCNC interactions involving 
the top quark~\cite{fritzsch,*peczha,*hanpeczha,*divpetsil,frihol},
which has a mass of the order of the electroweak energy scale, offer a
potentially new view of physics beyond the SM.

The FCNC-induced couplings of the type $tuV$ or $tcV$ (with
$V=\gamma,\z0$) have been explored in $\pp$ collisions at
the Tevatron by searching for the top-quark decays
$t\rightarrow uV$ and $t\rightarrow cV$~\cite{Tevatron}. The same couplings
involving the top quark were investigated in $\ele$
interactions at LEP2 by searching for single-top production through
the reactions $\ele\rightarrow t\bar u\ (+{\rm c.c.})$ and 
$\ele\rightarrow t\bar c\ (+{\rm c.c.})$~\cite{aleph,*aleph2,*opal,l3a}.
No evidence for such interactions was found at either accelerator
and limits were set on the branching ratios 
\mbox{$B(t\rightarrow q\gamma)$} and \mbox{$B(t\rightarrow qZ)$}.

\begin{figure}[p]
\vfill
\setlength{\unitlength}{1.0cm}
\begin{picture} (18.0,8.0)
\put (2.5,-2.0){\epsfig{figure=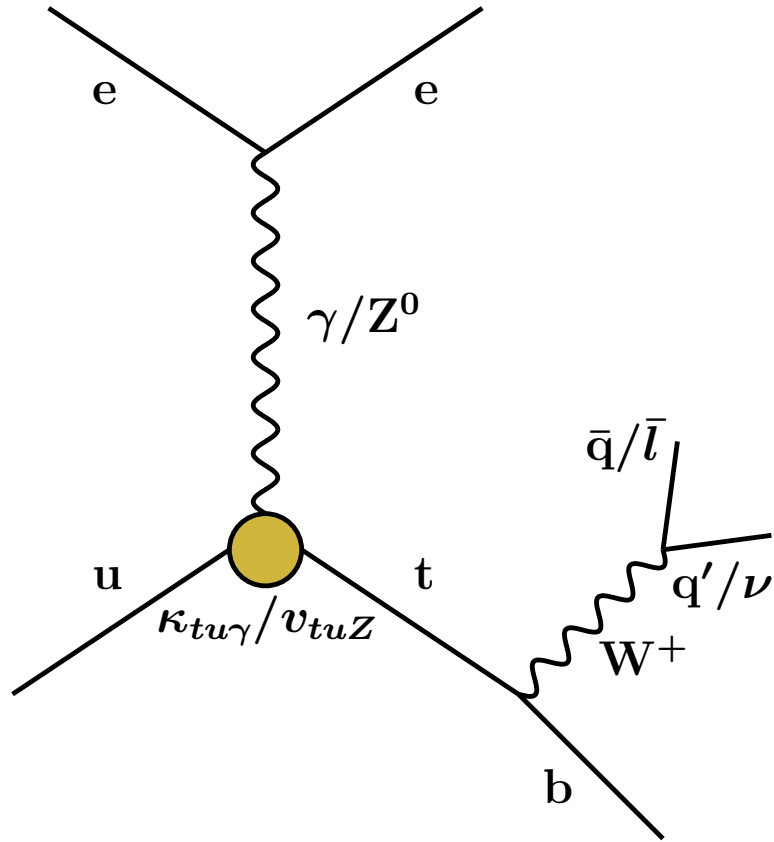,width=12cm}}
\end{picture}
\caption{
Single-top production via flavour-changing neutral current
  transitions at HERA.}
\label{one}
\vfill
\end{figure}

In $ep$ collisions at the HERA collider, top quarks can only be singly
produced. In the SM, single-top production proceeds through the
charged current (CC) reaction 
$ep\rightarrow \nu t\bar b {\rm X}$~\cite{schuler,*baurbij,*bijold}. 
Since the SM cross section at HERA is less than
$1$~fb~\cite{stelzer,*moretti}, any observed 
single-top event in the present data can be attributed to physics
beyond the SM.
The FCNC couplings, $tuV$ or $tcV$, would induce the neutral current
(NC) reaction $\ept$~\cite{frihol,belyaev}, in which the incoming lepton
exchanges a $\gamma$ or $Z$ with an up-type quark in the proton, yielding
a top quark in the final state. Due to the large $Z$ mass, this process
is most sensitive to a coupling of the type $tq\gamma$. Furthermore,
large values of $x$, the fraction of the proton momentum carried by the
struck quark, are needed to produce a top. Since the $u$-quark
parton distribution function (PDF) of the proton is dominant at large
$x$, the production of single top
quarks is most sensitive to a coupling of the type $tu\gamma$ (see
Fig.~\ref{one}).

\section{Theoretical framework}
\label{thefram}
Deviations from the SM predictions due to FCNC transitions involving
the top quark can be parameterised in terms of couplings of the type $tuV$
(with $V=\gamma,\z0$) and described by an effective Lagrangian
of the form~\cite{hanhew}
\begin{equation}
\Delta{\cal L}_{\rm eff} = e\ e_{t}\ \bar t\ \frac{i \sigma_{\mu\nu} q^{\nu}}{\Lambda}\ \kg\ u\ A^{\mu} + \frac{g}{2\cos\theta_W}\ \bar t\ \gamma_{\mu}\ \vz\ u\ Z^{\mu}\ + {\rm h.c.},
\end{equation}
where $e$ ($e_t$) is the electron (top-quark) electric charge,
$g$ is the weak coupling constant, $\theta_W$ is the weak mixing angle, 
$\sigma_{\mu\nu}=\frac{1}{2}(\gamma^{\mu}\gamma^{\nu}-\gamma^{\nu}\gamma^{\mu})$,
$\Lambda$ is an effective cutoff which, by convention, is set to the
mass of the top quark, $\mt$, taken as 175 GeV, $q$ is the momentum of
the gauge boson and $A^{\mu}$ ($Z^{\mu}$) is the photon ($Z$)
field. In the following,
it was assumed that the magnetic coupling $\kg$ and the vector coupling
$\vz$ are real and positive. The values of $\kg$ and $\vz$ in the SM
are zero at tree level and extremely small at the one-loop level.

The cross section for the process $\ept$ was calculated as a function of
$\kg$ including next-to-leading-order (NLO) QCD corrections in the eikonal
approximation~\cite{belyaev}. The renormalisation ($\mu_R$) and
factorisation ($\mu_F$) scales were chosen to be
$\mu_R=\mu_F=\mt$. The strong coupling constant,
$\alpha_s$, was calculated at two loops with
$\Lambda^{(5)}_{\overline{\rm MS}}=220$~MeV, corresponding to
$\asz=0.1175$. The calculations were performed using the
MRST99~\cite{epj:c4:463,*epj:c14:133} parameterisations of the proton
PDFs. The uncertainty of the results due to terms beyond NLO,
estimated by varying $\mu_R=\mu_F$ between $\mt/2$ and $2\mt$,
was $^{+1.6}_{-3.8}\%$ ($^{+1.3}_{-3.6}\%$) at a centre-of-mass energy of
318 (300)~GeV. The uncertainties of the results due to that on
$\asz$ and on the proton PDFs were $\pm 2\%$ and $\pm 4\%$,
respectively. The variation of the cross section on $\mt$ was
approximately $\pm 20\%$ ($\pm 25\%$) for $\Delta \mt=\pm 5$~GeV at a
centre-of-mass energy of 318 (300)~GeV.

\section{Experimental conditions}
\label{secsetup}
The data samples were collected with the ZEUS detector at HERA and
correspond to an integrated luminosity of 
$47.9\pm 0.9\ (65.5\pm 1.5)$~\pb1\ for $e^+p$ collisions taken during
1994-1997 (1999-2000) and $16.7\pm 0.3$~\pb1\ for $e^-p$ collisions
taken during 1998-99. During 1994-97 (1998-2000), HERA operated with
protons of energy $E_p=820$~GeV ($920$~GeV) and positrons or electrons
of energy $E_e=27.5$~GeV, yielding a centre-of-mass energy of
$\sqrt s=300$~GeV ($318$~GeV).

The ZEUS detector is described in detail 
elsewhere~\cite{zeus:1993:bluebook,pl:b293:465}. The main components
used in the present analysis were the central tracking detector
(CTD)~\cite{nim:a279:290,*npps:b32:181,*nim:a338:254}, positioned in a
1.43~T solenoidal magnetic field, and the uranium-scintillator sampling
calorimeter (CAL)~\cite{nim:a309:77,*nim:a309:101,*nim:a321:356,*nim:a336:23}. 

Tracking information is provided by the CTD, in which the momenta of
tracks in the polar-angle\footnote{The ZEUS coordinate system is a
right-handed Cartesian system, with the $Z$ axis pointing in the proton
beam direction, referred to as the ``forward direction'', and the $X$
axis pointing left towards the centre of HERA. The coordinate origin is
at the nominal interaction point.} region $15^\circ < \theta < 164^\circ$
are reconstructed. The CTD consists of 72 cylindrical drift chamber
layers, organised in nine superlayers. The relative transverse momentum,
$p_T$, resolution for full-length tracks can be parameterised as 
$\sigma(p_T)/p_T=0.0058\ p_T \oplus 0.0065\oplus 0.0014/p_T$, with
$p_T$ in GeV.

The CAL covers $99.7\%$ of the total solid angle. It is divided into
three parts with a corresponding division in $\theta$, as viewed
from the nominal interaction point: forward (FCAL,
$2.6^{\circ}<\theta<36.7^{\circ}$), barrel (BCAL,
$36.7^{\circ}<\theta<129.1^{\circ}$), and rear (RCAL,
$129.1^{\circ}<\theta<176.2^{\circ}$). Each of the CAL parts
is subdivided into towers which in turn are segmented longitudinally into
one electromagnetic (EMC) and one (RCAL) or two (FCAL, BCAL) hadronic
(HAC) sections. The smallest subdivision of the CAL is called a cell.
Under test-beam conditions, the CAL single-particle energy resolution is
$\sigma(E)/E=18\%/\sqrt{E}$ for electrons
and $\sigma(E)/E=35\%/\sqrt{E}$ for hadrons, with $E$ in GeV.

The luminosity was measured using the Bethe-Heitler reaction 
$ep\rightarrow e\gamma p$. The resulting small-angle energetic photons
were measured by the luminosity
monitor~\cite{desy-92-066,*zfp:c63:391,*acpp:b32:2025}, a lead-scintillator
calorimeter placed in the HERA tunnel at $Z=-107$~m.

\subsection{Trigger conditions}
\label{trigcon}
A three-level trigger was used to select events
online~\cite{zeus:1993:bluebook,proc:chep:1992:222}.
At the first level,
events were selected using criteria based on either the transverse
energy or missing transverse momentum measured in the CAL.
Events were accepted with a low threshold on these quantities when a
coincidence with CTD tracks from the event vertex was required, while
a higher threshold was used for events with no CTD tracks.

At the second level, timing information from the CAL was used to reject
events inconsistent with an $ep$ interaction. In addition, the
topology of the CAL energy deposits was used to reject non-$ep$ background
events. Cuts on the missing transverse momentum of $6$~GeV ($9$~GeV for
events without CTD tracks) or on the total transverse energy of $8$~GeV,
excluding the eight CAL towers immediately surrounding the forward
beampipe, were applied.

At the third level, track reconstruction and vertex finding were
performed and used to reject events with a vertex inconsistent with
the distribution of $ep$ interactions. Events with missing transverse
momentum in excess of $7$~GeV or containing at least two jets with
transverse energy $\etjet>6$~GeV and pseudorapidity $\etajet<2.5$ were
accepted; the latter condition was based upon the application of a
jet-finding cone algorithm with radius $R=1$ applied to the CAL cell
energies and positions.

\section{Monte Carlo simulation}
\label{secmc}
Samples of events were generated using Monte Carlo (MC)
simulations to determine the selection efficiency for the signal of
single-top production through FCNC processes and to estimate
background rates from SM processes. The generated events were passed
through the GEANT~3.13-based~\cite{tech:cern-dd-ee-84-1} ZEUS
detector- and trigger-simulation programs
\cite{zeus:1993:bluebook}. They were reconstructed and analysed by the
same program chain as the data.

Single-top production through FCNC processes in $ep$ collisions was
simulated using the HEXF generator~\cite{hexf}. Samples of events were
generated assuming top-quark masses of $170$, $175$ and $180$~GeV.
Initial-state radiation
from the lepton beam was included using the Weizs\"acker-Williams
approximation~\cite{wwa}. The hadronic final state was simulated using
the matrix-element and parton-shower model of LEPTO~\cite{cpc:101:108}
for the QCD cascade and the Lund string model~\cite{prep:97:31} as
implemented in JETSET~\cite{cpc:39:347,*cpc:43:367} for the
hadronisation. The MRSA~\cite{pr:d50:6734} parameterisations of the
proton PDFs were used.

The most important background to the positron-decay channel of the $W$
in the chain $t\rightarrow bW^+\rightarrow be^+\nu$ arose from NC
deep inelastic scattering (DIS). Two-photon processes provide a source of
high-$p_T$ leptons that were a significant background to the
muon-decay channel of the $W$ in the chain 
$t\rightarrow bW^+\rightarrow b\mu^+\nu$. In addition, single-$W$
production was a significant source of background to 
$t\rightarrow bW^+$, in both the positron- and muon-decay channels of the
$W$. The dominant
source of background for the hadronic-decay channel of $W$ in the chain
$t\rightarrow bW^+\rightarrow bq\bar q^{\prime}$ was multi-jet
production from QCD processes.

Several MC programs were used to simulate the different background
processes. The NC DIS events were generated using the LEPTO~6.5
program~\cite{cpc:101:108} interfaced to HERACLES~4.6.1~\cite{heracles,*spi:www:heracles}
via DJANGOH~1.1~\cite{cpc:81:381,*spi:www:djangoh11}. The HERACLES program includes photon
and $Z$ exchanges and first-order electroweak radiative
corrections. The QCD cascade was modelled with the colour-dipole
model~\cite{pl:b165:147,*pl:b175:453,*np:b306:746,*zfp:c43:625} by
using the ARIADNE~4.08 program~\cite{cpc:71:15,*zfp:c65:285} and
including the boson-gluon-fusion process. As an alternative, samples
of events were generated using the model of LEPTO based on first-order
QCD matrix elements plus parton showers (MEPS). In both cases, the
hadronisation was performed using the Lund string model. The
CTEQ5D~\cite{cteq5} parameterisations for the proton PDFs
were used. Two-photon processes were simulated using the generator
GRAPE~1.1~\cite{grape}, which includes dilepton production via
$\gamma\gamma$, $Z\gamma$ and $ZZ$ processes and considers both
elastic and inelastic production at the proton vertex. Single-$W$
production was simulated using the event generator EPVEC~\cite{epvec},
which did not include hard QCD radiation. Recent cross-section
calculations including higher-order QCD
corrections~\cite{jp:g25:1434,*hep-ph-9905469,*peer,*hep-ex-0302040} and using the
CTEQ4M~\cite{cteq4} (ACFGP~\cite{acfgp}) proton (photon) PDFs
were used to reweight the EPVEC event samples. Multi-jet QCD
production at low $\q2$, where $\q2$ is the virtuality of the
exchanged photon, was simulated using
PYTHIA~5.7~\cite{cpc:46:43,*cpc:82:74}. In this generator, the
partonic processes were simulated using leading-order (LO) matrix
elements, with the inclusion of initial- and final-state parton
showers. Hadronisation was performed using the Lund string model. The
MRSA (GRV-HO~\cite{pr:d46:1973}) parameterisations of the proton
(photon) PDFs were used.

\section{Signatures of FCNC-induced single-top production}
\label{secsel}
Single-top production via the FCNC coupling at the $tu\gamma$
vertex in $ep$ collisions at HERA, $\ept$, is predicted to proceed
predominantly through the exchange of a quasi-real photon between the beam
electron or positron and a valence $u$ quark in the proton (see
Fig.~\ref{one}). According to the signal MC simulation, the scattered
electron or positron escapes through the rear beampipe, outside the CAL
acceptance, in $65\%$ of the events.

In this analysis, the top-quark search was optimised for the decay
$t\rightarrow bW^+$. In the leptonic decay channel of the $W$, the
signal for such events is the presence of 
an isolated high-energy lepton, significant missing transverse
momentum arising from the emitted neutrino and a jet stemming from the
$b$-quark decay. 
In the hadronic decay channel of the $W$, the signal is the presence
of three jets in the final state with the dijet invariant-mass
distribution for the correct pair of jets peaking at the mass of the
$W$ boson, $\mw$, and the three-jet invariant-mass distribution peaking
at $\mt$.

\section{Leptonic channel}
\subsection{Data selection}
\label{lepsel}
Events with isolated high-energy leptons ($e^{\pm}$ or $\mu^{\pm}$),
significant missing transverse momentum and a jet were selected.
Similar previous analyses have been done by the H1~\cite{epj:c5:575} and
ZEUS~\cite{pl:b471:411} Collaborations.
Positron candidates were identified using an algorithm that
combined CAL and CTD information~\cite{zfp:c74:207}.
Muons were identified by the coincidence of a track in the CTD with
significant transverse momentum and CAL energy deposits consistent
with those expected from a minimum ionizing particle.
The charge information on the candidates was not used and they are
generically referred to as positrons and muons. The main selection
criteria are:
\begin{itemize}
\item cuts on the CAL timing and $Z$ coordinate ($|Z|<50$~cm) of the
  event vertex and algorithms based on the pattern of tracks in the
  CTD were used to reject events not originating from $ep$ collisions;
\item the track associated with the positron or muon
  candidate was required to have $p_T^{\rm track}>5$~GeV. To reduce
  the NC DIS background, the track was required to have
  $\theta<115^{\circ}$. In addition, it must have passed
  through at least three radial superlayers of the CTD (corresponding
  to $\theta\gtrsim 17^{\circ}$) and be isolated. Two
  isolation variables were defined for a given track using the
  separation $R$ in the $\etaphi$ plane, where
  $R=\sqrt{(\Delta\eta)^2+(\Delta\varphi)^2}$. The variable $D_{\rm
    jet}$ was defined as the distance from the nearest jet axis, while
  $D_{\rm track}$ was the distance from the nearest neighbouring track
  in the event. Events containing tracks with $D_{\rm jet}>1$ and
  $D_{\rm track}>0.5$ were selected;
\item $p_T^{\rm CAL}>20$~GeV, where $p_T^{\rm CAL}$ is the missing
  transverse momentum as measured with the CAL. It was reconstructed
  using the energy deposited in the CAL cells, after corrections for
  non-uniformity and dead material located in front of the
  CAL~\cite{epj:c11:427}. Energy deposits originating from identified
  muons were excluded from the measurement of $p_T^{\rm CAL}$;
\item the presence of at least one jet with transverse energy $\etjet$
  above $5$~GeV and $-1<\etajet<2.5$ was required. The longitudinally
  invariant $\kt$ cluster algorithm~\cite{np:b406:187} was used in the
  inclusive mode~\cite{pr:d48:3160} to reconstruct jets from the
  energy deposits in the CAL cells. The jet search was performed in
  the $\etaphi$ plane of the laboratory frame. The axis of each jet
  was defined according to the Snowmass
  convention~\cite{proc:snowmass:1990:134}, where $\etajet$
  ($\phijet$) was the transverse-energy-weighted mean pseudorapidity
  (azimuth) of all the cells belonging to that jet. The jet transverse
  energy was reconstructed as the sum of the transverse
  energies of the cells belonging to the jet and was corrected for
  detector effects such as energy losses in the inactive material in
  front of the CAL~\cite{pl:b531:9}. In the leptonic channel, only
  those jets for which the electromagnetic-energy fraction was
  below 0.9 and $R_{90\%}\geq 0.1$, where $R_{90\%}$ is the radius of
  the cone in the $\etaphi$ plane concentric to the jet axis that
  contains $90\%$ of the jet energy, were considered;
\item in events with an identified positron candidate, the
  acoplanarity angle, $\Phi_{\rm ACOP}$, was defined as the azimuthal
  separation of the outgoing positron and the vector in the
  $(X,Y)$-plane that balances the hadronic system. For well measured
  NC DIS events, the acoplanarity angle is close to zero, while a large
  $\Phi_{\rm ACOP}$ indicates large missing energy, as expected
  from top-quark decays. To reduce the background from NC DIS
  processes, the acoplanarity angle was required to be greater than
  $8^{\circ}$.
\end{itemize}

The selected data sample contained 36 events, 24 of which had a positron
candidate and 12 a muon candidate (see Table~\ref{tabsec1}).

\subsection{Comparison with Monte Carlo simulations}
\label{lepmc}
\begin{figure}[p]
\vfill
\setlength{\unitlength}{1.0cm}
\begin{picture} (18.0,20.5)
\put (4.5,15.0){\epsfig{figure=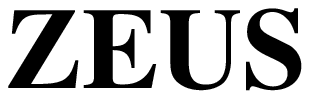,width=8cm}}
\put (1.5,14.5){\epsfig{figure=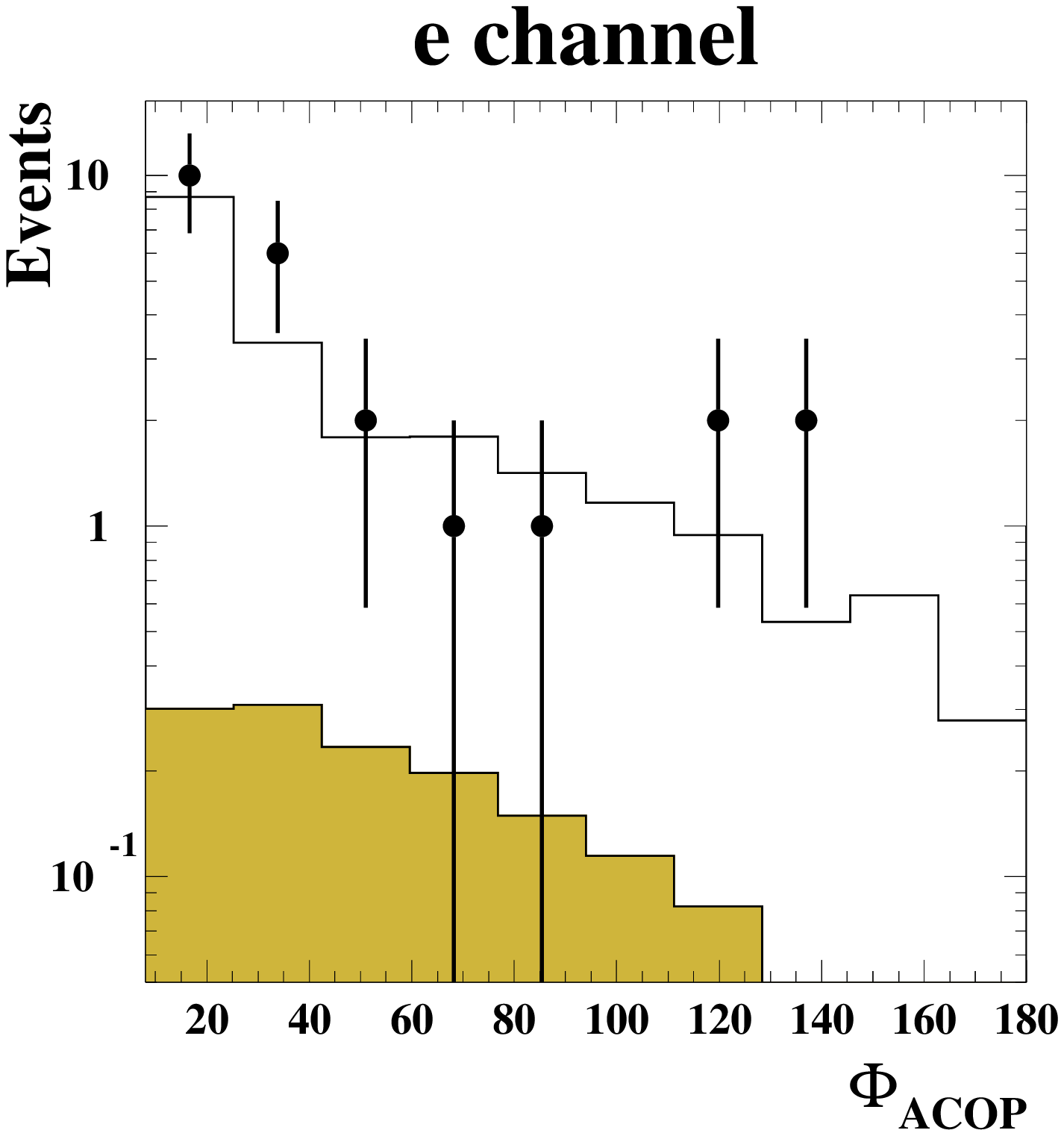,width=8cm}}
\put (7.5,14.5){\epsfig{figure=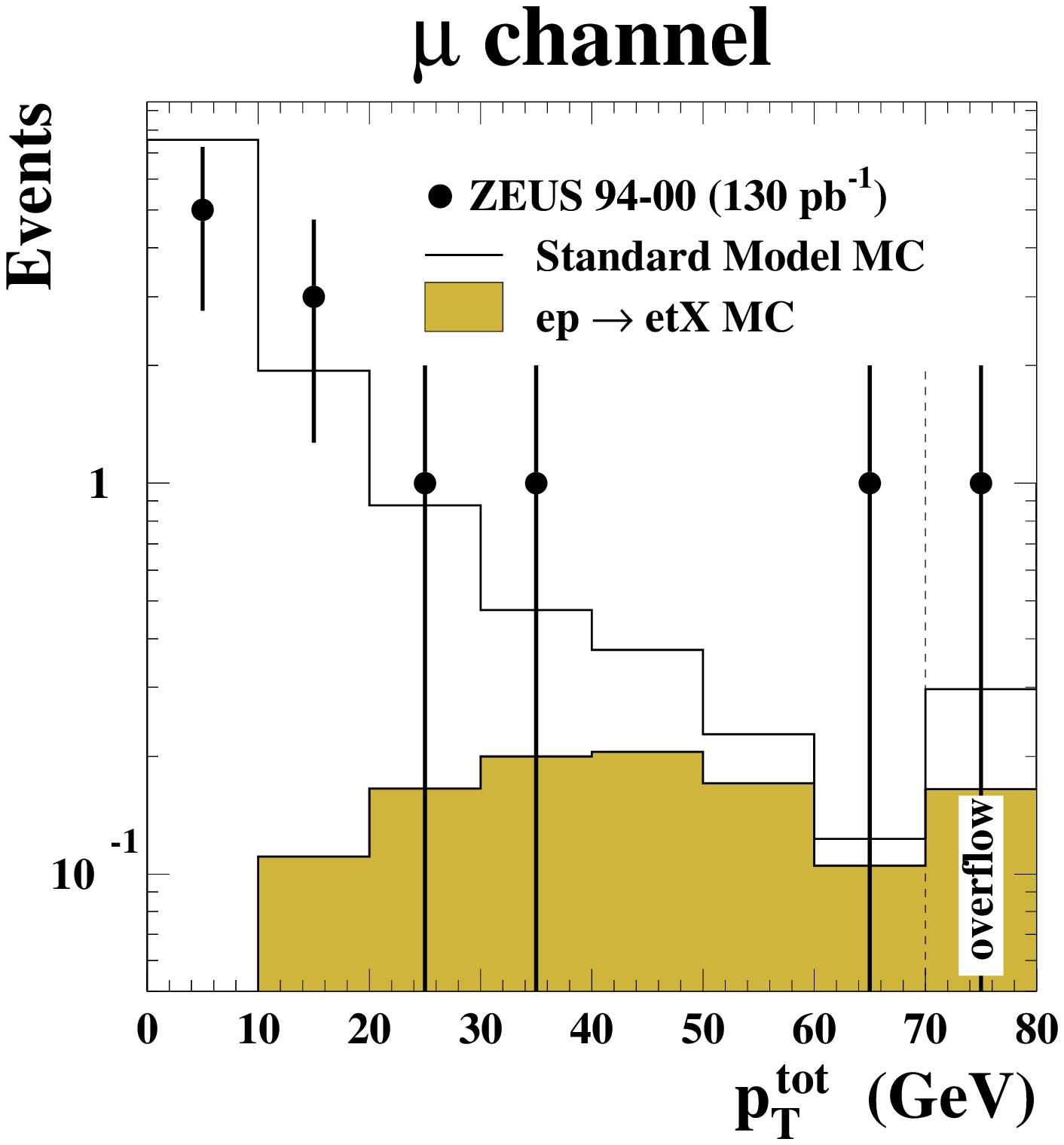,width=8cm}}
\put (1.5,8.3){\epsfig{figure=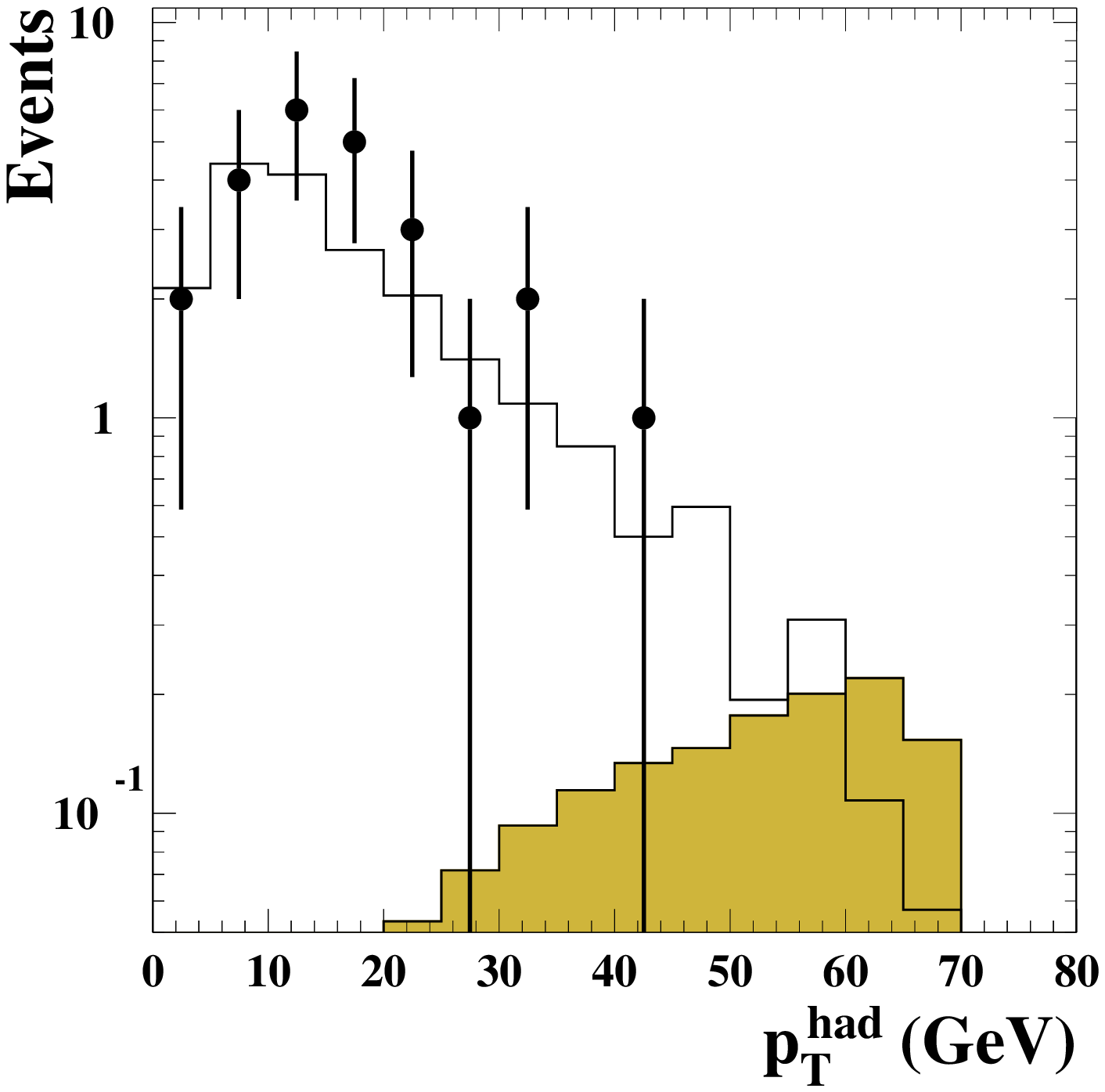,width=8cm}}
\put (7.5,8.3){\epsfig{figure=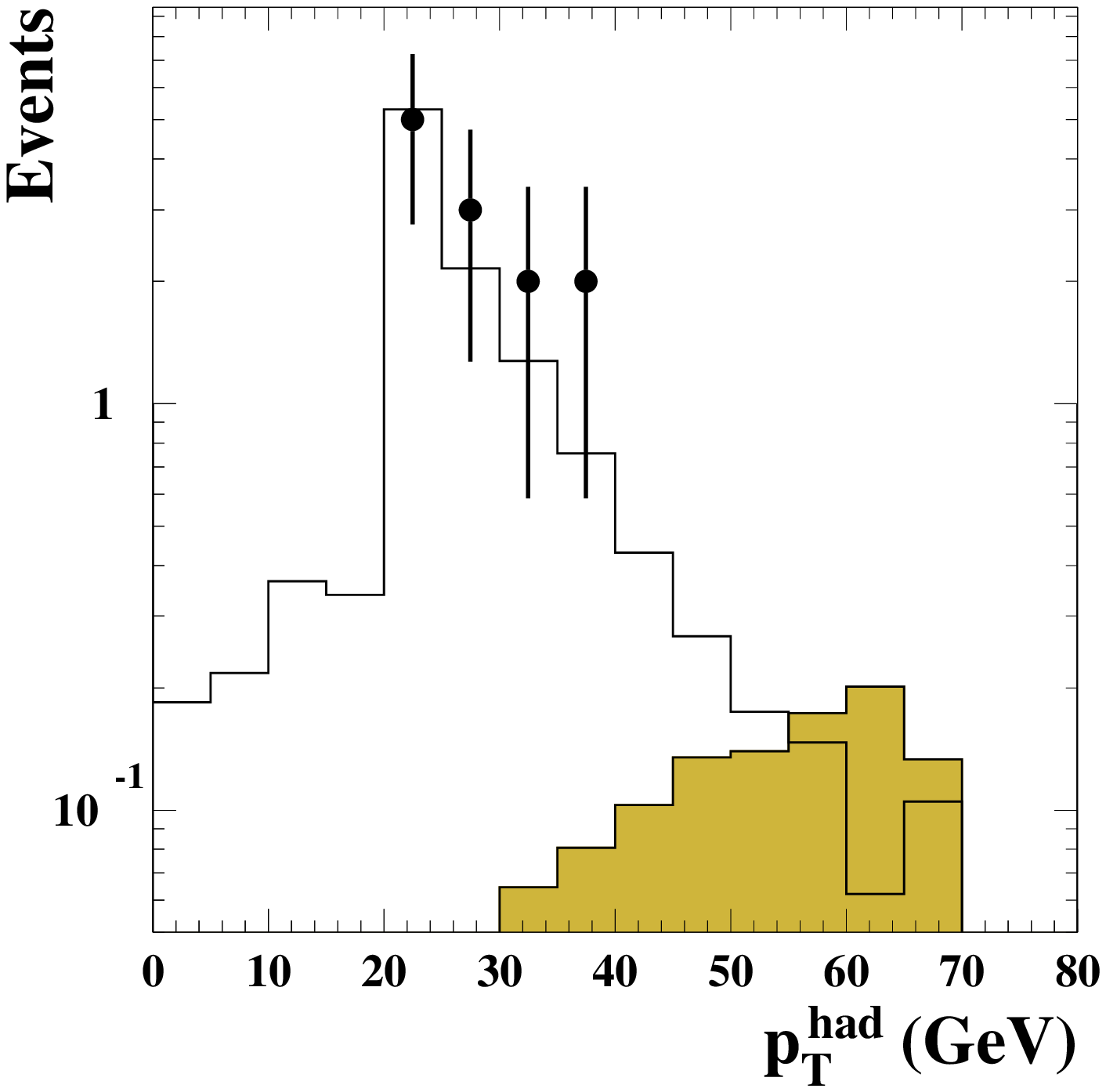,width=8cm}}
\put (1.5,2.0){\epsfig{figure=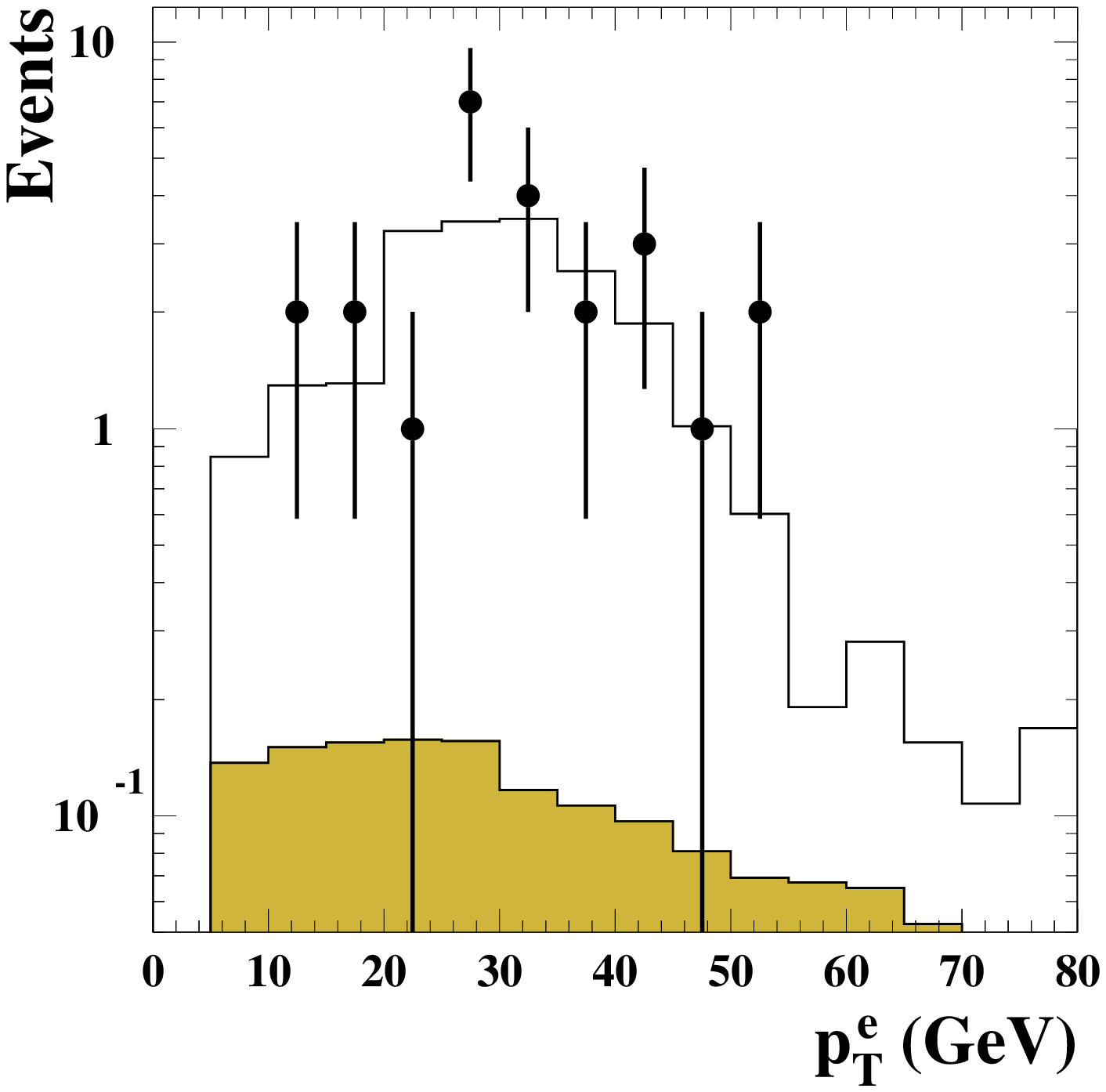,width=8cm}}
\put (7.5,2.0){\epsfig{figure=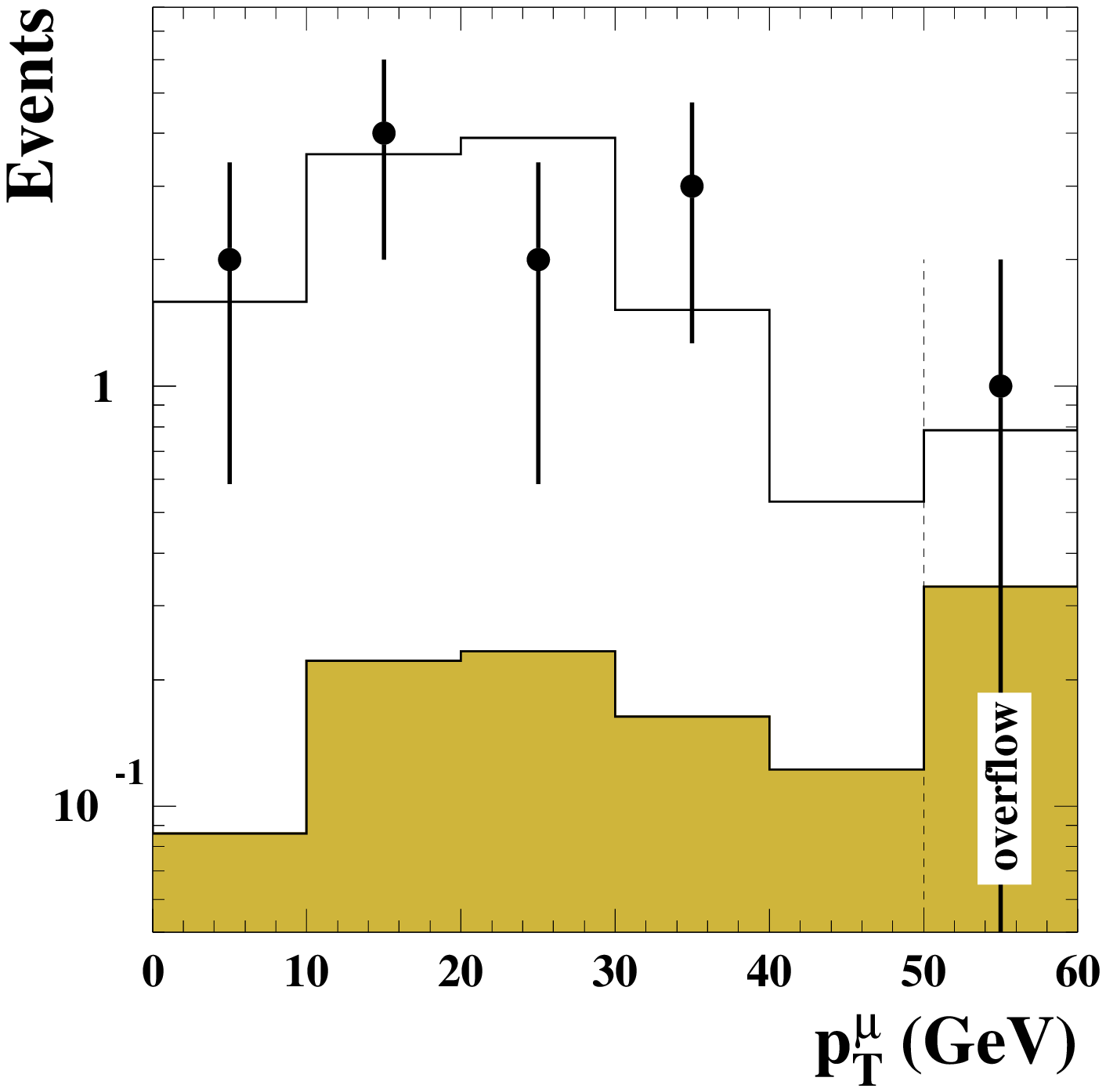,width=8cm}}
\put (7.3,20.3){\bf\small a)}
\put (13.3,20.5){\bf\small d)}
\put (7.3,14.1){\bf\small b)}
\put (13.3,14.1){\bf\small e)}
\put (7.3,7.8){\bf\small c)}
\put (13.3,7.8){\bf\small f)}
\end{picture}
\vspace{-3.5cm}
\caption{
a) $\Phi_{\rm ACOP}$, b) $p_T^{\rm had}$, and c) $p_T^e$ for those
events with an identified positron candidate. d) $p_T^{\rm tot}$, e) 
$p_T^{\rm had}$ and f) $p_T^{\mu}$ for those events with an identified
muon candidate. The dots are the data, the solid histogram is the
Standard Model MC simulation and the shaded histogram represents the
signal with $\mt=175$~GeV normalised to the limit presented in
Section~\ref{results}. The final bins in d) and f), marked
``overflow'', contain all events above the lower boundaries of these
bins. The distributions are for the selected events according to the
criteria of Section~\ref{lepsel}. The Standard Model MC distributions
have been normalised to the luminosity of the data.
}
\label{two}
\vfill
\end{figure}

The properties of the selected events were studied in detail and
compared with the MC predictions of the SM. Figures~\ref{two}a)-c)
show the acoplanarity, the transverse momentum of the hadronic system,
$p_T^{\rm had}$, and the transverse momentum of the positron
candidate as measured in the CAL, $p_T^e$, for those events with an
identified positron candidate. Figures~\ref{two}d)-f) show the
CAL transverse momentum corrected for the muon momentum
measured by the CTD, 
$p_T^{\rm tot}=\sqrt{(p_X^{\rm CAL}+p_X^{\mu})^2+(p_Y^{\rm CAL}+p_Y^{\mu})^2}$,
$p_T^{\rm had}$ and the transverse momentum of the track associated
with the muon candidate, $p_T^{\mu}$, for events with an identified
muon candidate. In each case, the distribution of data events can be
accounted for by the simulation of SM processes. The SM expectation
for the positron channel is dominated by NC DIS and that for the muon
channel by two-photon processes.

\subsection{Results of the search in the leptonic channel}
\label{lepres}
The final event selection for the $bl\nu$ final state required a
high-$p_T$ jet and missing energy. The cuts were optimised
using the simulations of both the SM background and the expected
single-top signal. The cuts used were:
\begin{itemize}
\item $p_T^{\rm had}> 40$~GeV for both the positron and muon decays;
\item $\delta=\sum_i (E_i-E_i \cos{\theta_i})=\sum_i (E-p_Z)_i< 47$~GeV
     for the positron decay,
     where the sum runs over all CAL energy deposits with
     corrected energy $E_i$ and polar angle $\theta_i$~\cite{epj:c11:427}.
     For fully contained NC DIS events, $\delta$ peaks at $55$~GeV, i.e.
     twice the lepton beam energy, which follows from energy-momentum
     conservation;
\item $p_T^{\rm tot}>10$~GeV for the muon decay.
\end{itemize}
After applying these requirements, no event remained in the data
sample. The efficiency for detecting single-top production in the
leptonic channel was $34\%$ for the positron decay and $33\%$ for
the muon decay. These efficiencies do not include the branching
ratio of the top-quark decay in the corresponding channel. 

In a recent study~\cite{hep-ex-0301030}, the H1 Collaboration has
reported an excess of events for $p_T^{\rm had}>25$~GeV. The number of
selected events in each channel with $p_T^{\rm had}>25$~GeV for the
present analysis is also listed in Table~\ref{tabsec1}. These results
are in agreement with the expectations from the SM.

\section{Hadronic channel}
The data used for this channel correspond to a slightly reduced
luminosity of $127.2$~\pb1.
\subsection{Data selection}
\label{jetsel}
The expected signature for the hadronic-decay channel of single-top
production through the FCNC $tu\gamma$ coupling is three jets with
large $\etjet$ and no significant missing transverse momentum. Since
it is expected that for the bulk of the events the scattered
positron escapes through the rear beam pipe, NC DIS events with 
$\q2\gtrsim 1$~\g2 were rejected. The data selection 
used similar criteria as reported in a previous
publication~\cite{pl:b531:9}.
Jets were found in the hadronic final state using the same algorithm as
described in Section~\ref{lepsel}. The main selection criteria are:
\begin{itemize}
\item cuts on the $Z$ coordinate ($-38<Z<32$~cm) of the event vertex,
  the number of tracks pointing to the vertex and the number of tracks
  compatible with an interaction upstream in the direction of the
  proton beam were used to reject events not originating from $ep$
  collisions;
\item the presence of at least three jets within the pseudorapidity
  range $\etar$ was required. The three highest-$\etjet$ jets in the
  event, ordered according to decreasing $\etjet$, were further
  required to satisfy $E_T^{\rm jet(1,2,3)}>40,\ 25,\ 14$~GeV;
\item CC DIS events were rejected by requiring the missing transverse
  momentum to be small compared to the total transverse energy,
  $E^{\rm tot}_T$, i.e.
  \mbox{$p_T^{\rm CAL}/\sqrt{E^{\rm tot}_T}<2\ \sqrt{\rm GeV}$};
\item NC DIS events with an identified scattered-positron
  candidate~\cite{nim:a365:508,*nim:a391:360} in the CAL were removed
  from the sample using the method described in an earlier
  publication~\cite{pl:b322:287};
\item $8.8 < \delta < 52.2$~GeV. The upper cut removed unidentified
  NC DIS events and the lower cut rejected proton beam-gas interactions.
\end{itemize}
The selected sample contained 348~events.

The invariant mass of jets $k$ and $l$ was determined using the
corrected jet transverse energies, as explained in Section~\ref{lepsel},
and jet angular variables according to the formula
$$\mj = \sqrt{2 E_T^{\rm jet,k} E_T^{\rm jet,l}
                [ \cosh{(\eta^{\rm jet,k}-\eta^{\rm jet,l})} -
                  \cos{(\varphi^{\rm jet,k}-\varphi^{\rm jet,l})}]}.$$
The three-jet invariant mass, $\m3j$, was reconstructed using the
formula
$$\m3j = \sqrt{\sum_{k<l} 2 E_T^{\rm jet,k} E_T^{\rm jet,l}
                [ \cosh{(\eta^{\rm jet,k}-\eta^{\rm jet,l})} -
                  \cos{(\varphi^{\rm jet,k}-\varphi^{\rm jet,l})}]},$$
where the sum runs over $k,l=1,2,3$. 
The average resolution in $\mj$ was $8\%$ for $\mj>50$~GeV and the 
distribution of $\mj$ for all pairs of jets in a sample of MC
signal events is shown in Fig.~\ref{three}a). The average resolution
in $\m3j$ was $4\%$ for $\m3j>80$~GeV and the $\m3j$ distribution in a
sample of MC signal events is shown in Fig.~\ref{three}b). Cuts on
$\mj$ and $\m3j$ were used to search for a signal of single-top
production in the hadronic channel.

\subsection{Comparison with Monte Carlo simulations}
\label{hadmc}
The properties of the selected events were studied in detail and
were compared with the MC predictions of the SM processes.
The MC distributions were normalised to the number of events in the
data with $\m3j<159$ GeV, i.e. outside the region where the signal for
single-top production is expected. The resulting normalisation factor
was $1.11\pm 0.08$, which can be attributed to higher-order QCD
corrections to the jet cross sections. The simulations of SM processes
provide a reasonable description of the $\etjet$ and $\etajet$ data
distributions (not shown). The distribution of $\mj$ for all pairs of
jets in an event and that of $\m3j$ are presented in
Figs.~\ref{three}a) and b), respectively. The $\m3j$ distribution
shows a steep fall-off from $\m3j \sim 130$~GeV to 240~GeV. The SM
simulations describe the observed $\mj$ and $\m3j$ distributions
reasonably well.

\subsection{Results of the search in the hadronic channel}
\label{hadres}
The MC simulations of the signal and SM processes were
used to find optimal windows in $\mj$ and $\m3j$ for the
observation of a signal relative to the background. The resulting
windows were $65.2<\mj<90.8$~GeV and $159<\m3j<188$~GeV.

\begin{figure}[p]
\vfill
\setlength{\unitlength}{1.0cm}
\begin{picture} (18.0,20.5)
\put (4.5,15.0){\epsfig{figure=stop0.eps,width=8cm}}
\put (1.5,14.5){\epsfig{figure=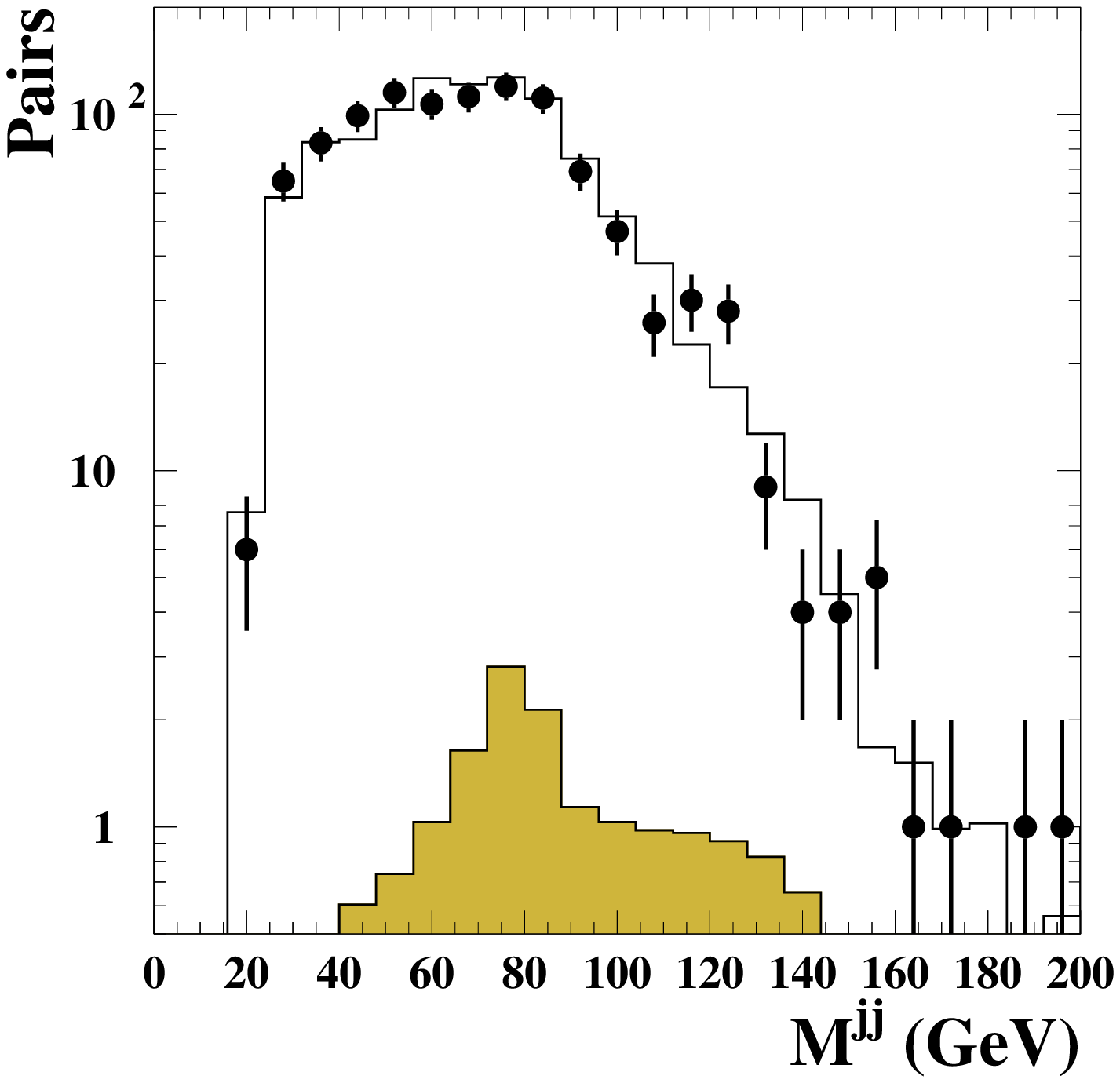,width=8cm}}
\put (7.5,14.5){\epsfig{figure=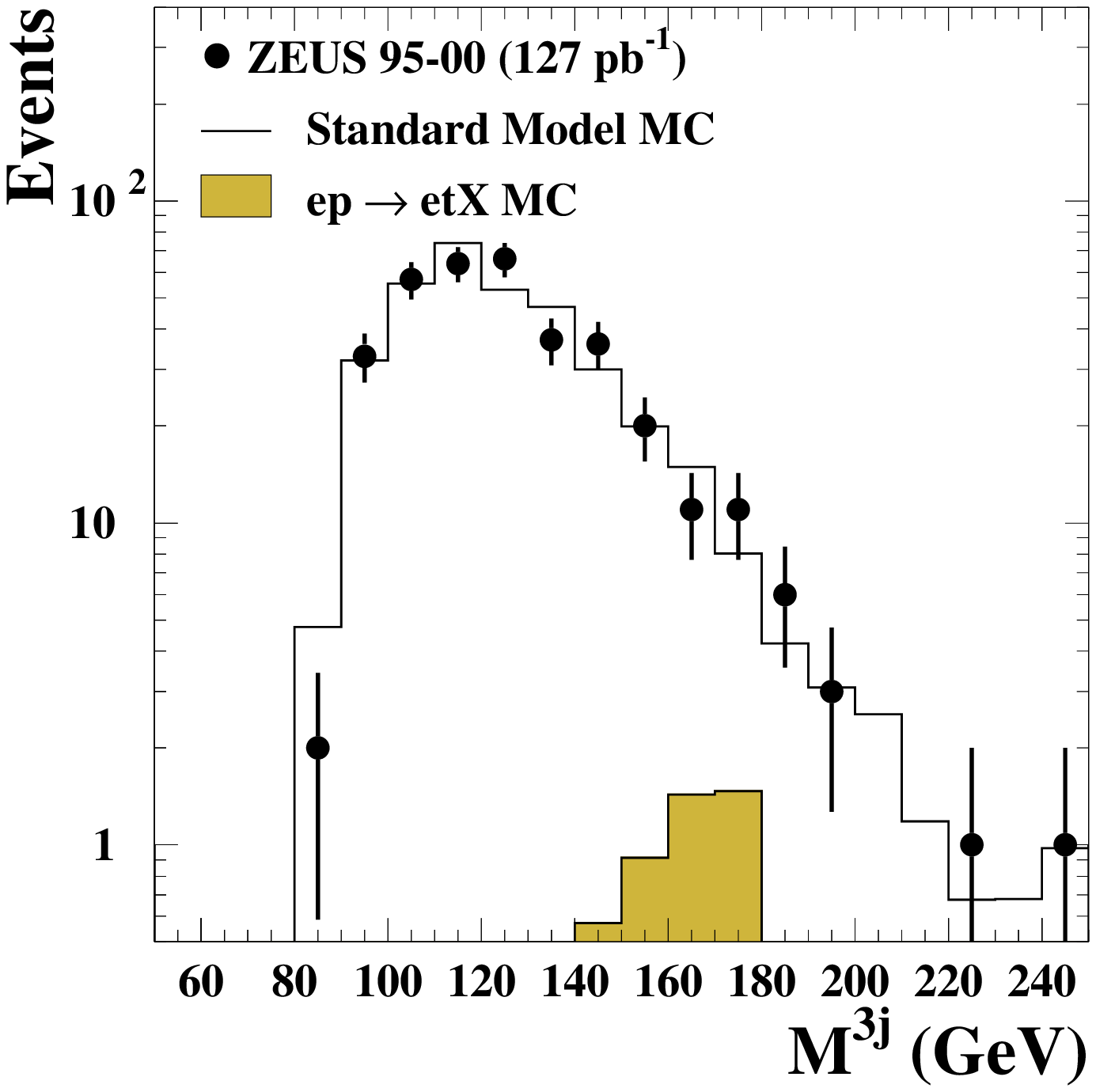,width=8cm}}
\put (1.5,8.3){\epsfig{figure=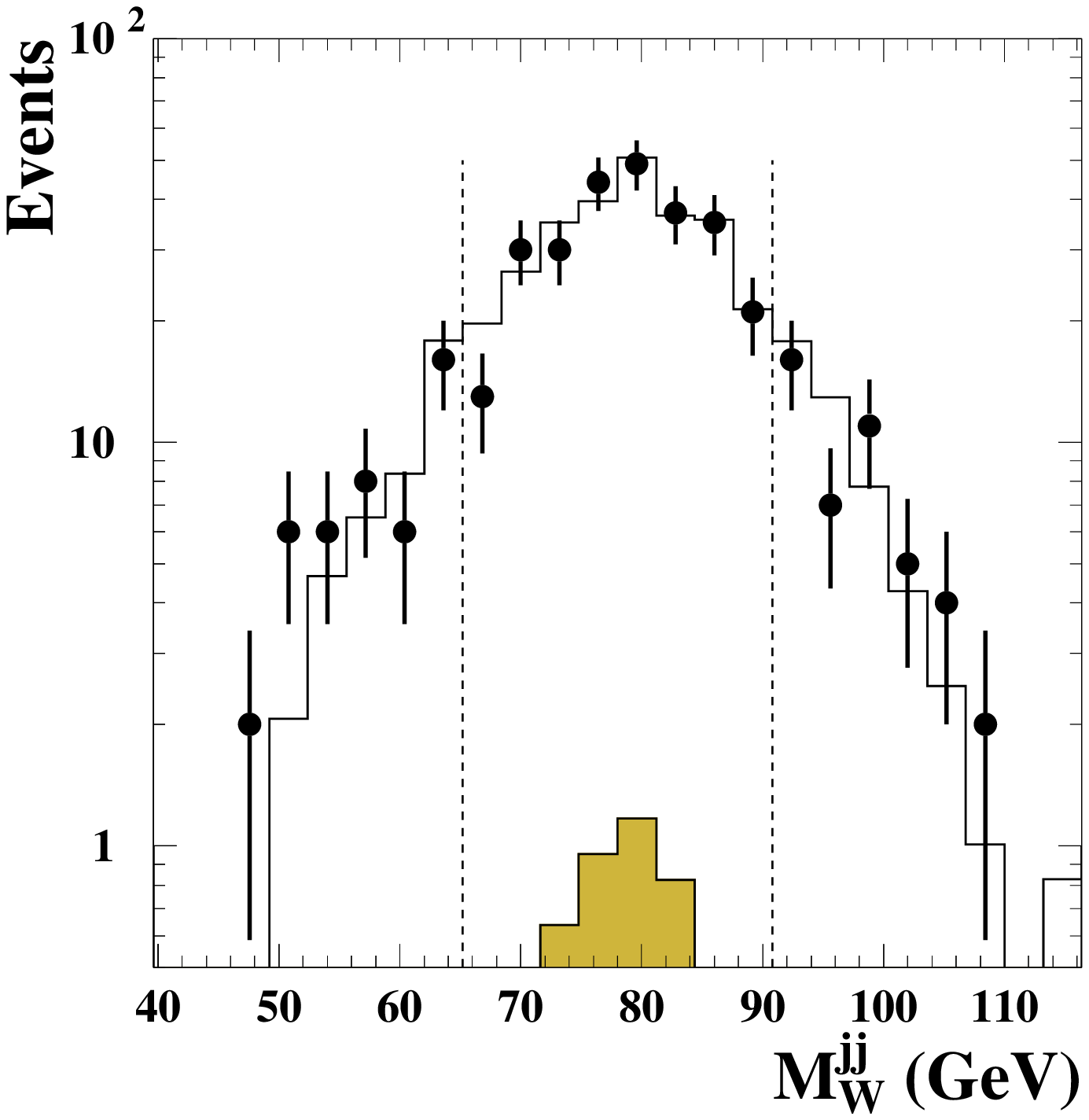,width=8cm}}
\put (7.5,8.3){\epsfig{figure=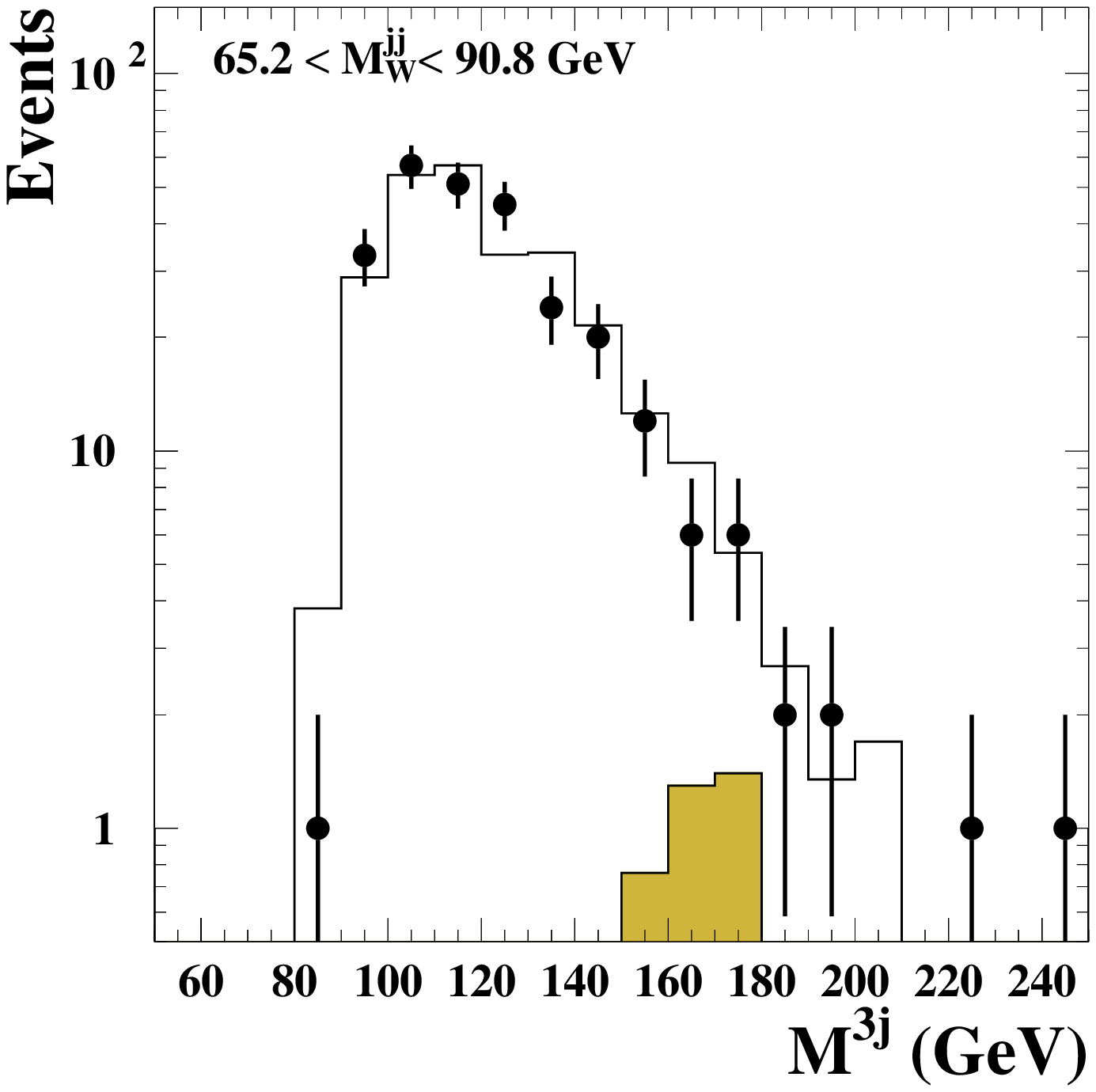,width=8cm}}
\put (1.5,2.0){\epsfig{figure=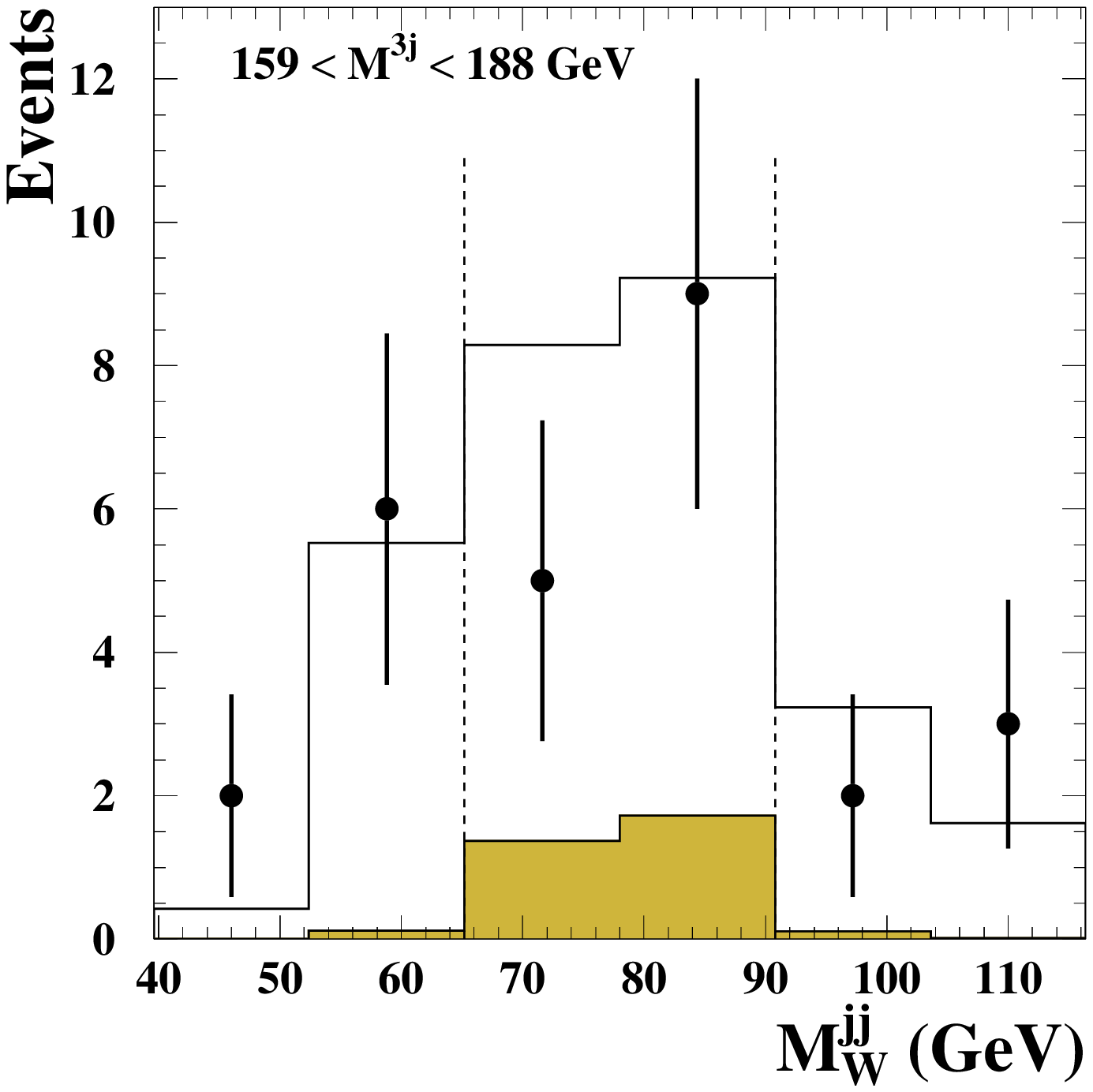,width=8cm}}
\put (7.5,2.0){\epsfig{figure=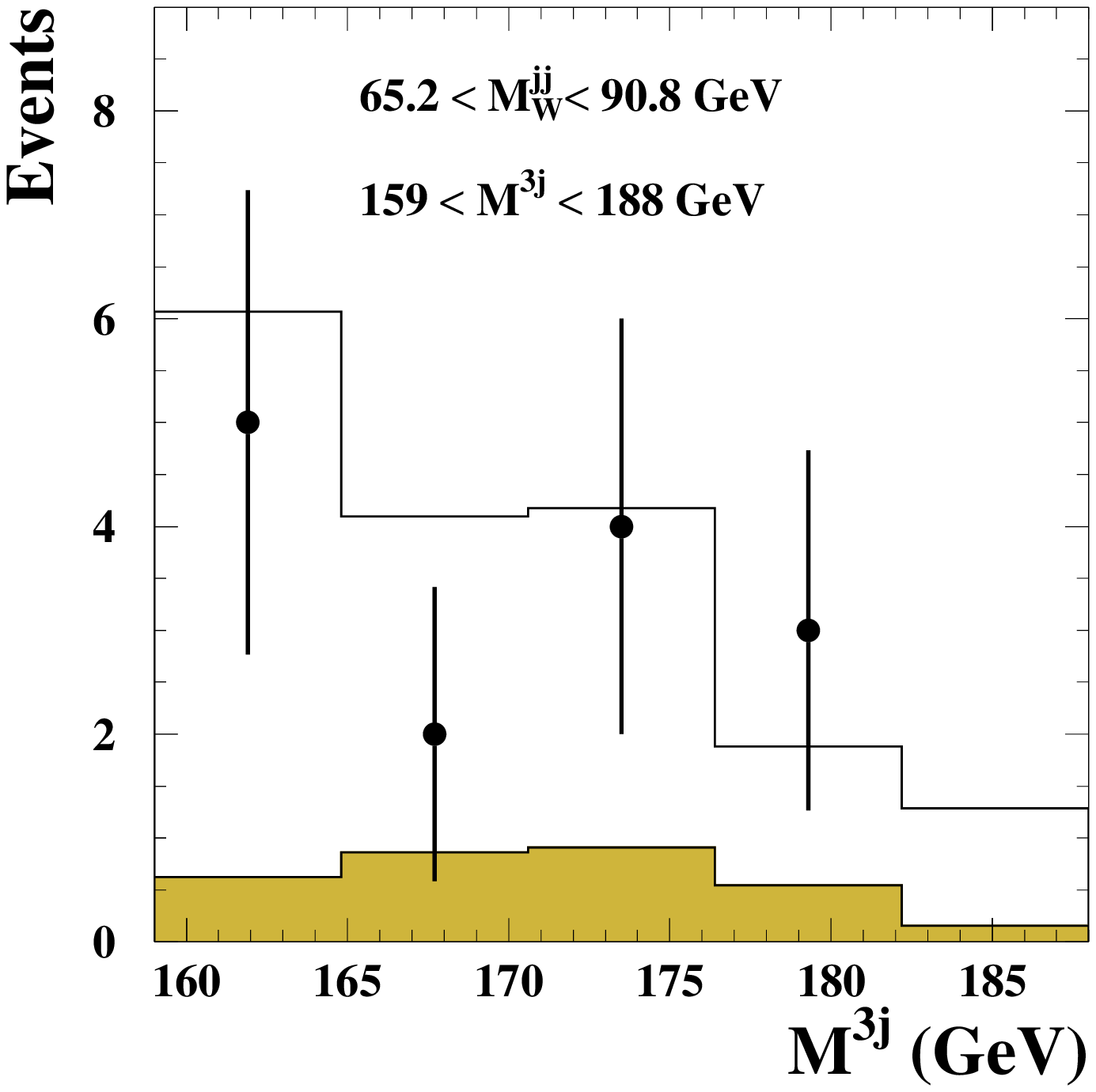,width=8cm}}
\put (7.3,20.3){\bf\small a)}
\put (13.3,20.3){\bf\small b)}
\put (7.3,14.1){\bf\small c)}
\put (13.3,14.1){\bf\small d)}
\put (7.3,7.8){\bf\small e)}
\put (13.3,7.8){\bf\small f)}
\end{picture}
\vspace{-3.5cm}
\caption{
  a) $\mj$ (for all pairs of jets), b) $\m3j$ and c) $M_W^{\rm jj}$
  distributions for the sample of events selected in the hadronic
  channel; d) $\m3j$ distribution for those events with
  $65.2<M_W^{\rm jj}< 90.8$~GeV; e) $M_W^{\rm jj}$ distribution for
  those events with $159<\m3j<188$~GeV; f) $\m3j$ distribution for
  those events with $65.2< M_W^{\rm jj}<90.8$~GeV. The dashed lines
  in c) and e) represent
  the cut at $65.2< M_W^{\rm jj}<90.8$~GeV. The distributions are for
  three-jet events with $E_T^{\rm jet1}>40$, $E_T^{\rm jet2}>25$,
  $E_T^{\rm jet3}>14$~GeV and $-1 <\etajet<2.5$. Other details are
  given in the caption to Fig.~\ref{two}.}
\label{three}
\vfill
\end{figure}

The $\mj$ closest to $M_W$
is denoted by $M_W^{\rm jj}$. The distribution of $M_W^{\rm jj}$ is
shown in Fig.~\ref{three}c); 261~events in the data
satisfied the condition $65.2<M_W^{\rm jj}<90.8$~GeV. The $\m3j$
distribution after this cut is shown in Fig.~\ref{three}d). The
simulation of SM processes reproduces the distributions well. After
the requirement $159<\m3j<188$~GeV, 14~events remained. The
distributions of $M_W^{\rm jj}$ and $\m3j$ in the data after this cut
are shown in Figs.~\ref{three}e) and f), respectively, and are well
reproduced by the simulation of SM processes. After these cuts, the
efficiency for detecting single-top production in the hadronic channel
was $24\%$. This efficiency does not include the branching ratio of
the top-quark decay in the hadronic channel. The observed $\m3j$
distribution shows no significant excess at $\mt$.

\section{Systematic uncertainties}
\label{systema}
The most important sources of systematic uncertainty were:
\begin{itemize}
\item leptonic channel\\
     $-$ the uncertainty of $\pm 1\%$ on the absolute energy scale of the
         CAL gave changes of $^{+6.7}_{-1.6}\%$ in the background and
         negligible changes in the signal-efficiency estimations;\\
     $-$ the use of the LEPTO-MEPS model instead of ARIADNE to
         estimate the NC DIS background gave a change of $-0.8\%$
         in the background estimation;\\
     $-$ the MC statistical uncertainty on the SM background
         estimation was $\pm 7.5\%$;\\
\item hadronic channel\\
     $-$ the uncertainty of $\pm 1\%$ on the absolute energy scale of the
         jets~\cite{pl:b531:9} gave changes of $^{+10.4}_{-1.7}\%$ in
         the background and $^{+3.9}_{-4.9}\%$ in the
         signal-efficiency estimations;\\
     $-$ the MC normalisation uncertainty on the SM background estimation
         was $\pm 7.4\%$.
\end{itemize}

All these uncertainties in the number of expected background events were
added in quadrature and are shown in Tables~\ref{tabsec1} and
\ref{tabsec2}. The experimental uncertainties were smaller than the
theoretical uncertainties and, therefore, were not considered in the
derivation of the limits for single-top production.

\section{Limit on the FCNC couplings}
\label{results}

As no event was selected in the leptonic channel and no excess over
the SM prediction was observed in the hadronic channel (see
Table~\ref{tabsec2}), limits were set on FCNC 
couplings of the type $tqV$. The contribution of the charm quark, which
has only a small density in the proton at high $x$, was ignored by setting
$\kappa_{tc\gamma}=v_{tcZ}=0$. Only the anomalous couplings involving
a $u$ quark, $\kg$ and $\vz$, were considered.

At HERA, most of the sensitivity to FCNC-induced couplings involving
the top quark comes from the process $\ept$ in which a $\gamma$ is
exchanged since the large $Z$ mass suppresses the contribution due to
$Z$ exchange. In a first step, limits on $\kg$ were, therefore, derived
assuming $\vz=0$ and using NLO QCD calculations of the cross section
for the process $\ept$ (see Section~\ref{thefram}). The results
obtained from each channel and centre-of-mass energy together with
those from the combined analysis presented below are summarised in
Table~\ref{tabsec2}. Limits from a combination of channels were
obtained by using a method described in a previous
publication~\cite{pr:d63:052002}. In the derivation of the limits, the
decrease in the branching ratio $B(t\rightarrow Wb)$ in the presence of
FCNC decays was taken into account. 
In comparison to the dependence of the result on the assumed value of
$\mt$, the effects of all other uncertainties are very
small. Therefore, limits were evaluated for $\mt=170,\ 175$ and $180$ GeV,
neglecting the other uncertainties.

By combining the results from both the leptonic and hadronic channels,
an upper limit of
$$\kg<0.174\ {\rm at}\ 95\%\ {\rm CL}$$
was derived assuming $\mt=175$~GeV. The limit was $\kg<0.158\ (0.210)$ for
$\mt=170\ (180)$~GeV. The above coupling limit corresponds to a
limit on the cross section for single-top
production of
$$\sigma(\ept,\sqrt s=318\ {\rm GeV}) <
0.225 \ {\rm pb}\ \ {\rm at} \ 95\% \ {\rm CL}.$$

In a second step, the effects of a non-zero $\vz$ coupling were taken into
account. The derivation of the exclusion region in the $\kg-\vz$ plane
was made using LO calculations for the process $\ept$ obtained with
the program CompHEP~\cite{comphep}, since NLO corrections
to the contribution from $Z$ exchange are not available.
Limits in the $\kg-\vz$ plane were derived by using a two-dimensional
probability density evaluated assuming a Bayesian prior probability
distribution flat in $\kg$ and $\vz$:
\begin{equation}
\rho(\kg,\vz |D)=\frac{\prod_i L_i(N^i_{\rm obs}|\kg,\vz)}{\int_0^{\infty}d\kg
 \int_0^{\infty} d\vz \ \prod_i L_i(N^i_{\rm obs}|\kg,\vz)}, \nonumber
\end{equation}
where $\rho(\kg,\vz |D)$ is the probability density for the FCNC couplings
given the set of observed data $D$ and $L_i(N^i_{\rm obs}|\kg,\vz)$
are the partial likelihoods for each channel and centre-of-mass energy
evaluated as the Poissonian probabilities to observe $N^i_{\rm obs}$
events given the expectations of the SM background processes and the
signals for single-top production. The $95\%$ CL limit was found as
the set of points $\rho(\kg,\vz|D)=\rho_0$ such that
$${\int\int}_{\rho(\kg,\vz|D)>\rho_0} d\kg\ d\vz \ \rho(\kg,\vz|D) = 0.95.$$

\begin{figure}[p]
\vfill
\setlength{\unitlength}{1.0cm}
\begin{picture} (18.0,9.0)
\put (-4.0,-5.0){\epsfig{figure=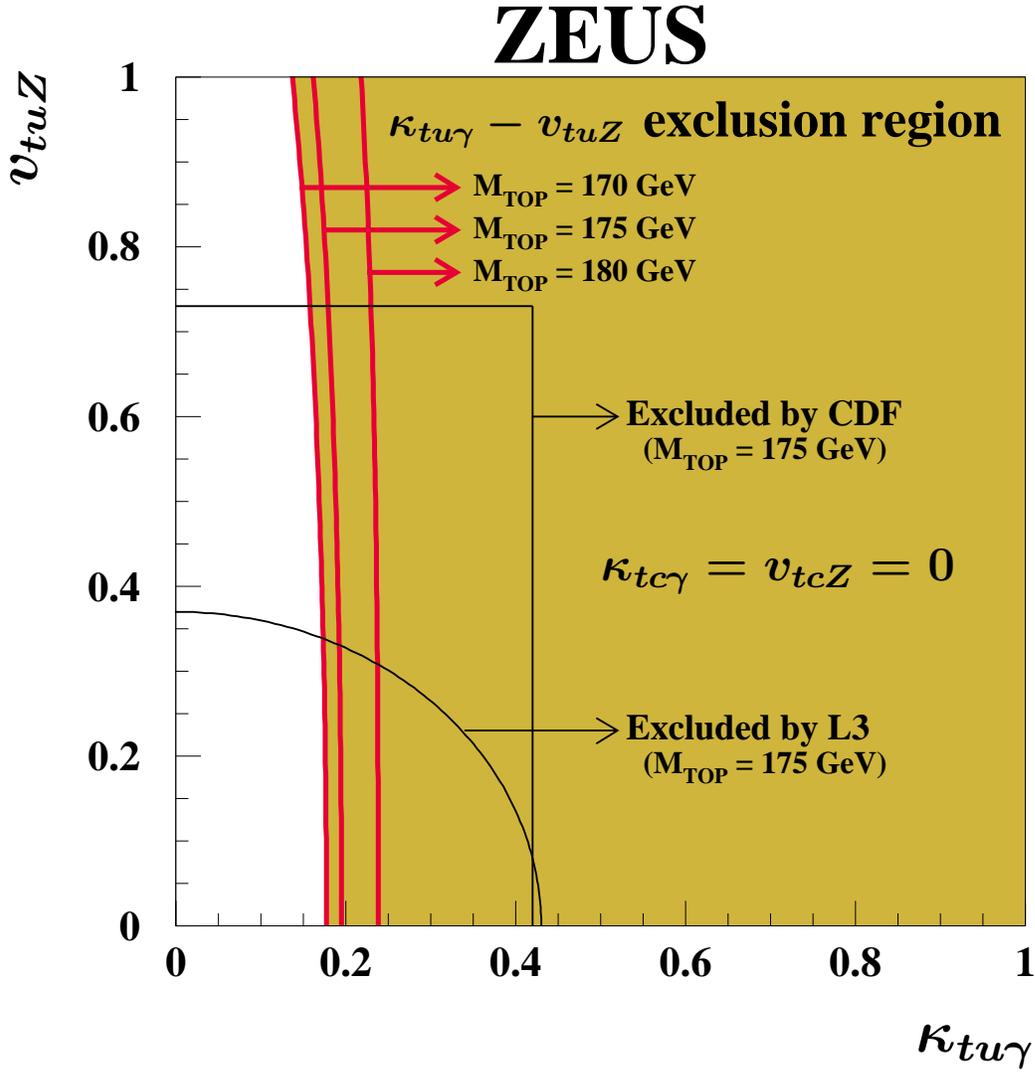,width=22cm}}
\end{picture}
\caption{
Exclusion regions at $95\%$ CL in the $\kg-\vz$ plane for three values
of $\mt$ (170, 175 and 180~GeV) assuming $\kappa_{tc\gamma}=v_{tcZ}=0$.
The CDF and L3 exclusion limits for $\mt=175$ GeV are also shown.}
\label{four}
\vfill
\end{figure}

Figure~\ref{four} shows the exclusion region on the $\kg-\vz$ plane
obtained from this search, together with those from
CDF~\cite{Tevatron,signature2} and L3~\cite{l3a}, which is the most stringent
limit from LEP2~\cite{aleph,*aleph2,*opal}. It should also be noted
that the Lagrangian used in the LEP
\mbox{analyses~\cite{aleph,*aleph2,*opal,l3a,signature2}} differs from that in Eq.~(1)
by a constant multiplicative factor such that
$\kappa_{tq\gamma}^{\rm LEP}=\sqrt{2}\ \kappa_{tq\gamma}^{\rm ZEUS}$ and
$v_{tqZ}^{\rm LEP}=\sqrt{2}\ v_{tqZ}^{\rm ZEUS}$.
In Fig.~\ref{four}, the limits from CDF and L3 are plotted using the
Lagrangian convention of Eq.~(1). The measurements at the Tevatron and
LEP have similar sensitivities to the $tuV$ and $tcV$ couplings, and
their limits were obtained with the assumptions $\kg=\kappa_{tcZ}$ and
$v_{tu\gamma}=v_{tcZ}$. In Fig.~\ref{four}, the published CDF and L3
limits are rescaled by $\sqrt 2$ for the purposes of comparison to the
present results on $\kg$ and $\vz$ which are obtained assuming
$\kappa_{tc\gamma}=v_{tcZ}=0$. The limit-setting procedure was
repeated assuming $\mt = 170$ and $180$~GeV; the resulting exclusion
regions are also shown in Fig.~\ref{four}.

\section{Summary}
\label{secsumm}
Single-top production via flavour-changing neutral current
transitions has been searched for with the ZEUS detector at HERA in
positron-proton and electron-proton collisions at centre-of-mass
energies of 300 and 318~GeV using an integrated luminosity of
$130.1$~\pb1.
No deviation from the Standard Model prediction was found. The results
were used to constrain single-top production $\ept$ via the FCNC
process. An upper limit on the FCNC coupling $\kg$ of $0.174$ at
$95\%$ CL was obtained. This limit excludes a substantial region in
$\kg$ not constrained by previous experiments.

\newpage
\noindent {\Large\bf Acknowledgements}
\vspace{0.3cm}

We thank the DESY Directorate for their strong support and encouragement.
The remarkable achievements of the HERA machine group were essential for
the successful completion of this work and are greatly appreciated. We
are grateful for the support of the DESY computing and network services.
The design, construction and installation of the ZEUS detector have been
made possible owing to the ingenuity and effort of many people from DESY
and home institutes who are not listed as authors.
A.~Belyaev and N.~Kidonakis provided the computer code to
make the cross-section calculations for single-top production. The NLO
calculations for $W$ production were provided by K.-P.~O.~Diener, 
C.~Schwanenberger and M.~Spira. In
addition, we wish to thank E. Boos for useful discussions.

\newpage
\clearpage
{\raggedright

\providecommand{\etal}{et al.\xspace}
\providecommand{\coll}{Collaboration}
\catcode`\@=11
\def\@bibitem#1{%
\ifmc@bstsupport
  \mc@iftail{#1}%
    {;\newline\ignorespaces}%
    {\ifmc@first\else.\fi\orig@bibitem{#1}}
  \mc@firstfalse
\else
  \mc@iftail{#1}%
    {\ignorespaces}%
    {\orig@bibitem{#1}}%
\fi}%
\catcode`\@=12

}

\newpage
\clearpage
\begin{table}[p]
\centering
\begin{tabular}{|l||c|c|} \hline
 & Positron & Muon \\
 \multicolumn{1}{|c||}{Leptonic channel}  & channel & channel \\ 
 & obs./expected ($W$) & obs./expected ($W$) \\ \hline
 \multicolumn{3}{c}{ } \\ \hline 
 \multicolumn{3}{|c|}{Preselection }   \\ \hline\hline
$e^+p$, $\sqrt s=300\ {\rm GeV}$ (${\cal L}=47.9$ \pb1) & 
4 / $7.3_{-2.1}^{+0.8}$ & 0 / $4.2_{-0.3}^{+0.4}$  \\ \hline
$e^-p$, $\sqrt s=318\ {\rm GeV}$ (${\cal L}=16.7$ \pb1) & 
7 / $3.2_{-1.0}^{+0.6}$ & 1 / $2.1_{-0.2}^{+0.2}$  \\ \hline
$e^+p$, $\sqrt s=318\ {\rm GeV}$ (${\cal L}=65.5$ \pb1) & 
13 / $10.1_{-1.9}^{+0.9}$ & 11 / $5.6_{-0.4}^{+0.4}$  \\ \hline\hline
Total (${\cal L}=130.1$ \pb1) & 
24 / $20.6_{-4.6}^{+1.7}\ (17\%)$ & 12 / $11.9_{-0.7}^{+0.6}\ (16\%)$ \\ \hline
 \multicolumn{3}{c}{ } \\ \hline 
 \multicolumn{3}{|c|}{Final selection ($p_T^{\rm had}>25$ GeV)}  \\ \hline\hline
$e^+p$, $\sqrt s=300\ {\rm GeV}$ (${\cal L}=47.9$ \pb1) & 
0 / $0.72_{-0.13}^{+0.27}$ & 0 / $0.78_{-0.10}^{+0.10}$ \\ \hline
$e^-p$, $\sqrt s=318\ {\rm GeV}$ (${\cal L}=16.7$ \pb1) & 
1 / $0.64_{-0.20}^{+0.28}$ & 1 / $0.45_{-0.07}^{+0.07}$ \\ \hline
$e^+p$, $\sqrt s=318\ {\rm GeV}$ (${\cal L}=65.5$ \pb1) & 
1 / $1.54_{-0.32}^{+0.33}$ & 4 / $1.53_{-0.16}^{+0.17}$ \\ \hline\hline
Total (${\cal L}=130.1$ \pb1) & 
2 / $2.90_{-0.32}^{+0.59}\ (45\%)$ & 5 / $2.75_{-0.21}^{+0.21}\ (50\%)$ \\ \hline
 \multicolumn{3}{c}{ } \\ \hline 
\multicolumn{3}{|c|}{Final selection ($p_T^{\rm had}>40$ GeV)}  \\ \hline\hline
$e^+p$, $\sqrt s=300\ {\rm GeV}$ (${\cal L}=47.9$ \pb1) & 
0 / $0.23_{-0.05}^{+0.05}$ & 0 / $0.26_{-0.04}^{+0.04}$ \\ \hline
$e^-p$, $\sqrt s=318\ {\rm GeV}$ (${\cal L}=16.7$ \pb1) & 
0 / $0.16_{-0.06}^{+0.06}$ & 0 / $0.08_{-0.01}^{+0.05}$ \\ \hline
$e^+p$, $\sqrt s=318\ {\rm GeV}$ (${\cal L}=65.5$ \pb1) & 
0 / $0.54_{-0.07}^{+0.07}$ & 0 / $0.61_{-0.09}^{+0.10}$ \\ \hline\hline
Total (${\cal L}=130.1$ \pb1) & 
0 / $0.94_{-0.10}^{+0.11}\ (61\%)$ & 0 / $0.95_{-0.10}^{+0.14}\ (61\%)$ \\ \hline
\end{tabular}
\caption{
Number of events in data and Standard Model
    background for the leptonic channel for different samples after
    the preselection and final selection cuts. The percentage of
    single-$W$ production included in the expectation is
    indicated in parentheses. The statistical and systematic uncertainties added
    in quadrature are also indicated.}
\label{tabsec1}
\end{table}

\begin{table}[p]
\centering
\begin{tabular}{|l||c|c||c|c|} \hline
 &
\multicolumn{2}{c||}{Leptonic channel} &
\multicolumn{2}{c|}{Hadronic channel} \\ \cline{2-5}
\multicolumn{1}{|r||}{$\sqrt s=$}
 & 300 GeV & 318 GeV & 300 GeV & 318 GeV \\ \hline\hline
$N_{\rm obs}$ & 0 & 0 & 5 & 9  \\ \hline
$N_{\rm SM}$ & $0.49_{-0.07}^{+0.07}$ & $1.40_{-0.13}^{+0.17}$ & $3.3_{-0.4}^{+1.3}$ & $14.3_{-1.1}^{+1.2}$ \\ \hline
$\epsilon\cdot Br$ ($\%$) & 6.9 & 7.1 & 16.6 & 16.5 \\ \hline
luminosity (\pb1) & 47.9 & 82.2 & 45.0 & 82.2 \\ \hline
$\sigma_{\rm lim}\times B(t\rightarrow Wb)$ (pb) & 0.906 & 0.514 & 0.998 & 0.426 \\ \hline
$\kg$ (per channel)            &
\multicolumn{2}{c||}{0.223} &
\multicolumn{2}{c|}{0.241} \\ \hline\hline
$\sigma_{\rm lim}$ (pb) (all channels) &
\multicolumn{4}{c|}{0.225 at $\sqrt s=318$ GeV} \\ \hline
$\kg$ (all channels)            &
\multicolumn{4}{c|}{0.174} \\ \hline
\end{tabular}
\caption{
Number of events in data and Standard Model
    background for the leptonic and hadronic channels for different
    samples, together with the efficiency times branching ratio of the
    signal and luminosity for each sample. The last four rows show the
    limits on the single-top production cross section via
    flavour-changing neutral current transitions and on the $\kg$
    coupling assuming $\mt=175$ GeV.}
\label{tabsec2}
\end{table}

\newpage
\clearpage

\newpage
\clearpage

\newpage
\clearpage

\newpage
\clearpage

\end{document}